# Room-temperature waveguide-integrated photodetector using bolometric effect for mid-infrared spectroscopy applications

Joonsup Shim[1], Jinha Lim[1], Inki Kim[1], Jaeyong Jeong[1], Bong Ho Kim[1], Seong Kwang Kim[1], Dae-Myeong Geum[2], and SangHyeon Kim[1*]

[1]School of Electrical Engineering, Korea Advanced Institute of Science and Technology (KAIST), 291 Daehak-Ro, Yuseong-Gu, Daejeon 34141, Republic of Korea.

[2]Department of Electrical & Computer Engineering, Inha University, 100, Inha-ro, Michuhol-gu, Incheon 22212, Republic of Korea.

*Corresponding Author: SangHyeon Kim (shkim.ee@kaist.ac.kr)

# Abstract

Waveguide-integrated mid-infrared (MIR) photodetectors are pivotal components for the development of molecular spectroscopy applications, leveraging mature photonic integrated circuit (PIC) technologies. Despite various strategies, critical challenges still remain in achieving broadband photoresponse, cooling-free operation, and large-scale complementary-metal-oxide-semiconductor (CMOS)-compatible manufacturability. To leap beyond these limitations, the bolometric effect – a thermal detection mechanism – is introduced into the waveguide platform. More importantly, we pursue a free-carrier absorption (FCA) process in germanium (Ge) to create an efficient light-absorbing medium, providing a pragmatic solution for full coverage of the MIR spectrum without incorporating exotic materials into CMOS. Here, we present an uncooled waveguide-integrated photodetector based on a Ge-on-insulator (Ge-OI) PIC architecture, which exploits the bolometric effect combined with FCA. Notably, our device exhibits a broadband responsivity of 28.35 %/mW across 4030–4360 nm (and potentially beyond), challenging the state of the art, while achieving a noise-equivalent power of $4.03 \times 10^{-7}$ W/Hz$^{0.5}$ at 4180 nm. We further demonstrate label-free sensing of gaseous carbon dioxide ($CO_2$) using our integrated photodetector and sensing waveguide on a single chip. This approach to room-temperature waveguide-integrated MIR photodetection, harnessing bolometry with FCA in Ge, not only facilitates the realization of fully integrated lab-on-a-chip systems with wavelength flexibility but also provides a blueprint for MIR PICs with CMOS-foundry-compatibility.

# Introduction

The mid-infrared (MIR) spectral region beyond 3 μm is of enormous scientific and technological importance, as it encompasses unique molecular fingerprints[1–3], enabling sophisticated chemical and biological analysis in a non-invasive manner through absorption spectroscopy techniques[4,5]. Leveraging highly mature photonic integrated circuit (PIC) technologies[6–8], substantial progress has been made in miniaturizing traditional external-optics-based spectrometers into chip-scale systems that offer cost-effective, mass-manufacturable, and scalable solutions[9–12]. A key bottleneck in realizing fully integrated and robust lab-on-a-chip systems is the monolithic integration of MIR photodetectors (PDs) into the waveguide platforms, which are an indispensable building block of PICs to convert light into electrical signals. Traditionally, surface-illuminated MIR PDs have relied on narrow-bandgap semiconductor materials such as HgCdTe alloy[13] and III-V compounds[14]; yet, they necessitate cryogenic cooling (bulky and costly) to mitigate high thermal noise at room temperature, posing severe challenges in practical applications[15]. More recently, two-dimensional materials (e.g., graphene[16] and black phosphorus[17]) have arisen as promising candidates that can be operated at room temperature[18]. However, the zero-bandgap nature of graphene results in an extremely high dark current level under biasing[19]. Additionally, black phosphorus itself exhibits severe performance degradation under ambient conditions and possesses an absorption edge of around 4 μm[20], restricting its utility to longer wavelengths. These emerging materials also still struggle with complementary-metal-oxide-semiconductor (CMOS)-compatible processes and wafer-scale integration, impeding large-scale and cost-saving production at commercialization levels[21].

In this regard, it is highly beneficial for MIR photodetection to harness thermal-type PDs, where the photoresponse is extracted by converting photo-induced heat generation into the electrical signal[18,22,23], enabling wavelength-insensitive photodetection by properly tailoring the spectral characteristics of the light absorber. Owing to the inherent nature of thermal detection, there is a slight compromise in response speed; however, a relatively moderate (or even low) level of bandwidth is adequate to meet the requirements for many MIR spectroscopy applications. To reap the full benefits of superior properties of the thermal detectors including wavelength independence and uncooled photodetection, here, we introduce the bolometric effect into the waveguide platform. Bolometers, a class of thermal-type PD, convert light-induced temperature fluctuations into changes in electrical resistance[23–25]. Notably, to date, few efforts have been devoted to achieving waveguide-integrated MIR photodetection beyond 3 μm using thermal detection mechanisms[26–29]. Optical absorbers with gold (Au) antennas on a suspended-Si waveguide have shown promise in bolometry in the range of 3.72–3.88 μm[26,27]; however, they offer low responsivities, and the resonance nature of the plasmonic structure intrinsically limits broadband photodetection. Additionally, the usage of noble metals is commonly restricted in CMOS foundries. Graphene-based detectors using the photothermoelectric (PTE) effect, another class of thermal detection mechanism based on the temperature-gradient-driven voltage generation governed by the Seebeck coefficient of the thermoelectric materials[30], have shown promising results on waveguide platforms such as chalcogenide glass (ChG)-on-CaF$_2$ at 5.2 μm[28] and Ge-on-Si (GOS) at 3.7 μm[29]. These graphene-based PTE detectors are particularly attractive due to their zero-bias operation and fast response times. However, their usage is limited to a wavelength range of ~8 μm due to the inherent material properties of the photonic platform, which restrict their utility across the broader MIR band. In addition, the use of plasmonic Au strips[29] and a split-gate architectures[28,29] introduces further fabrication challenges, including misalignment and errors during transfer and patterning of graphene. A more comprehensive comparison between bolometric and PTE detectors is provided in the *Discussion* section.

In this work, we remarkably advance the state of the art in waveguide-integrated MIR PDs by exploiting the bolometric effect with free-carrier absorption (FCA) in Ge[31,32] and titanium oxide-based bolometric material[33–35], providing a pragmatic approach with high fabrication robustness for uncooled MIR photodetection without foreign materials in CMOS or hybrid integrations. Our demonstration is based upon a CMOS-compatible Ge-on-insulator (Ge-OI) photonic platform with a buried oxide (BOX) of $Y_2O_3$ and a Si substrate, providing broad transparency window up to around 13 μm[36–40], and reaches record-high photoresponsivity for waveguide-integrated PDs using bolometric effect beyond 3 μm. Furthermore, to demonstrate the feasibility of non-destructive, label-free detection of molecules using our room-temperature-operated waveguide-integrated MIR PD, we have experimentally conducted gaseous carbon dioxide ($CO_2$) sensing by integrating the sensing waveguide and detector on a single chip.

## Results

**Device architecture and design.** Figure 1a illustrates a schematic of the proposed MIR PIC-based sensor on the Ge-OI platform, comprising a slot waveguide for analyte sensing (passive sensing part) and a waveguide-integrated PD (detector part), monolithically integrated onto a single chip. Our on-chip photonic sensor utilizes light–analyte interaction within the sensing waveguide through absorption spectroscopy[38,40] based on the Beer-Lambert law. An air-clad slot waveguide, supporting hollow-core guiding, has been employed to induce stronger light absorption with enhanced field confinement compared to conventional strip or rib waveguides, thus aiming to improve the sensitivity factor or to reduce the physical length of the sensing waveguide[11,36,38]. The residual light is then directly coupled from the sensing waveguide into the waveguide-integrated PD. As noted earlier, the operational principle of our proposed detector is the bolometric effect combined with FCA in Ge. For the bolometric material that converts light-induced temperature variations into changes in electrical resistance, we employed a $TiO_2/Ti/TiO_2$ tri-layer film, whose temperature-dependent electrical properties can be finely tailored by engineering the thickness of each layer[33–35]. The temperature change in bolometric detectors, in response to periodically varying incident light, can be described by[23]

$$\Delta T = \frac{\eta \Phi_0}{\sqrt{G_{th}^2 + \omega^2 \cdot C_{th}^2}} \quad (1)$$

where $\Delta T$ represents the temperature change, $\eta$ is the absorption efficiency for the given wavelength, $\omega$ and $\Phi_0$ are the angular frequency and the amplitude of the periodic radiation, respectively, $G_{th}$ is the thermal conductance between the detector and the surrounding environment, and $C_{th}$ is the thermal capacitance of the detector. As inferred from Equation (1), increasing $\eta$, while diminishing $G_{th}$ and $C_{th}$, is critical for enhancing $\Delta T$ for a given incident optical power, which directly correlates with the bolometric detector's responsivity. In order to boost $\eta$ within our PD, FCA in Ge should be elevated, which greatly depends on the type of free carriers and the doping concentration for particular wavelengths. To take full advantage of FCA in Ge, we selected heavily-doped p-type Ge ($p^+$ Ge) as the MIR-absorbing medium (details can be found in Supplementary Note 1). For the reduction of $G_{th}$ and $C_{th}$, optimizing device geometries is crucial. Here, the optimization process, including the geometrical parameters of the bolometer region – specifically, a length ($L_B$) of 4 μm and a width ($W_B$) of 8 μm – was conducted by considering heating efficiency, back-reflection, and in-house fabrication capabilities. The systematic process of optimizing geometries with numerical simulations is detailed in Supplementary Note 3. Figure 1b shows the simulated steady-state temperature distribution for the device designed with the final parameters. The input waveguide, having a width ($W_{in}$) of 2 μm, was designed to support only the fundamental transverse-electric (TE) mode. The incoming light was set to an optical power of 1 mW at a wavelength ($\lambda$) of 4.18 μm. As depicted in Fig. 1b, there is a significant temperature rise confined within the bolometer region. This localized heat generation is achieved by FCA within the $p^+$ Ge, which demonstrates the viability of an FCA-based thermalization process acting as a compact and efficient MIR absorber, even in the absence of resonance structures.

Figures 1c and 1d show the optical microscope and cross-sectional transmission electron microscopy (TEM) images, respectively, of the fabricated device on the Ge-OI photonic platform featuring a 500 nm-thick top Ge, a 2 μm-thick $Y_2O_3$ BOX, and a Si substrate. Here, the proposed waveguide-integrated PD incorporates a boron-doped $p^+$ Ge (bolometer region), a $SiO_2/Al_2O_3$ (20/25 nm) insulating layer stack, a bolometric material of $TiO_2/Ti/TiO_2$ (25/2/25 nm) tri-layer film, and a Ti/W (100/150 nm) metal electrode. The thickness of each layer in the bolometric material was carefully optimized (discussed in Supplementary Fig. S7). Additional characterizations, such as X-ray photoelectron spectroscopy (XPS) and X-ray diffraction (XRD), are detailed in Supplementary Figs. S9 and S10, respectively. Energy-dispersive X-ray spectroscopy (EDS) elemental mapping (see Supplementary Fig. S11) confirms the successful deposition of each layer. Secondary-ion mass spectrometry (SIMS) depth profile analysis in Fig. 1e quantitatively reveals the impurity dopant concentration within the $p^+$ Ge region. To make full use of FCA in Ge, ion implantation was performed with a high dopant dose of $5\times10^{15}$ cm$^{-2}$, and the implant energy was carefully adjusted to 110 keV. This optimization contributes to exposing a larger fraction of the modal field to the peak doping-concentration region of the absorbing medium, aligning with the mode-field maximum and the projected range ($R_p$) of the implanted ions in the $p^+$ Ge.

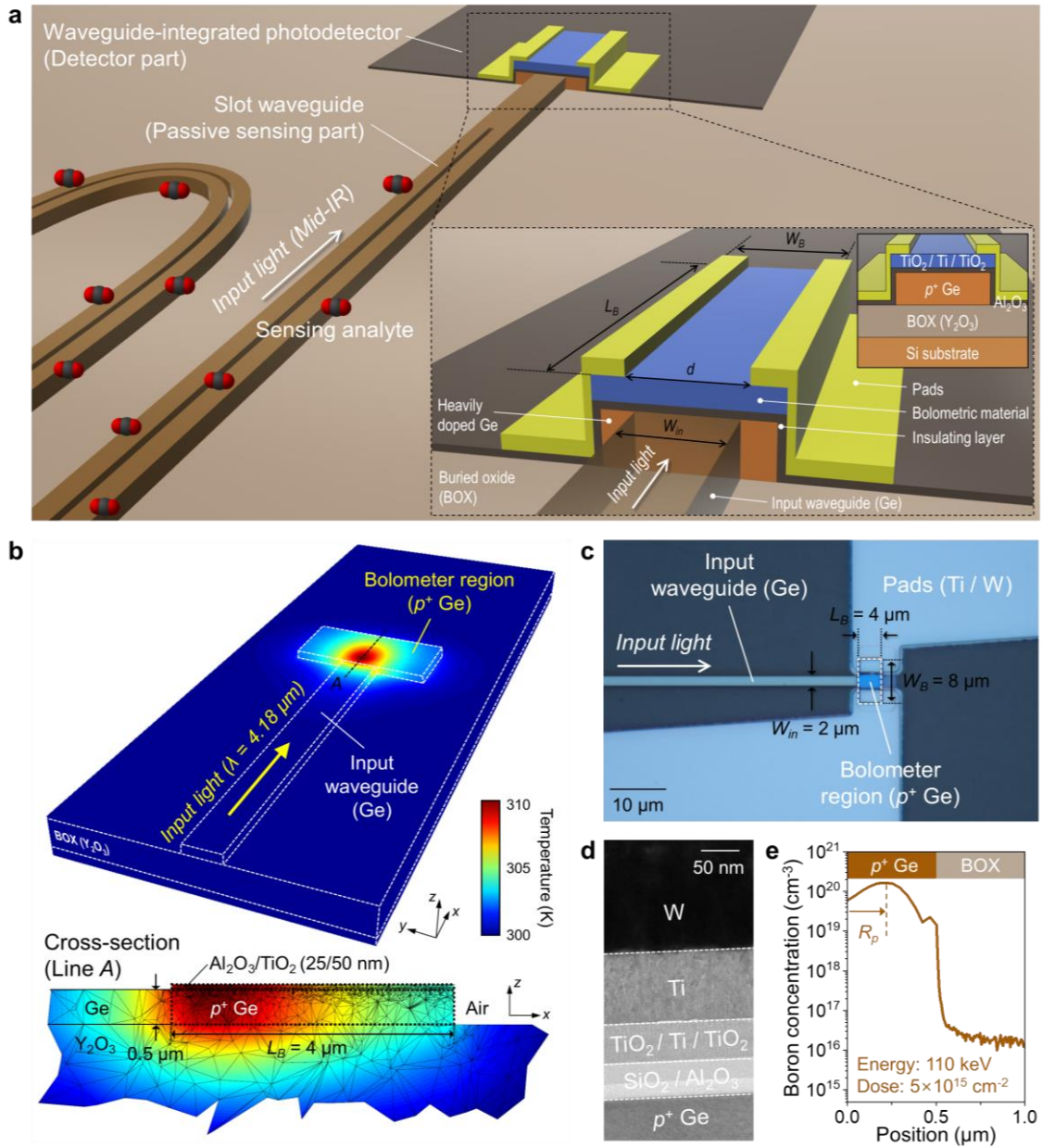

**Figure 1. Structure of the proposed device. a** Illustration of the MIR PIC-based sensor on the Ge-OI platform consisting of the sensing waveguide and photodetector. The zoom-in view shows the schematic of the proposed waveguide-integrated photodetector using the bolometric effect combined with FCA in Ge. **b** Heat simulation (Ansys Lumerical) results showing steady-state temperature distribution within the proposed photodetector ($W_B$ = 8 μm, $L_B$ = 4 μm) for an incoming light at 4.18 μm (1 mW). Doping concentration of $p^+$ Ge region was assumed as $10^{20}$ cm$^{-3}$. Background temperature was set to be 300 K. **c** Optical microscope image of the fabricated device. **d** Cross-sectional TEM image of the device including the electrode region. **e** SIMS depth profile analysis indicating the boron-doped $p^+$ Ge region. The implant energy and dopant dose were 110 keV and 5×10$^{15}$ cm$^{-2}$, respectively.

**Thermo-electrical characterization.** We first investigated the temperature-dependent electrical properties. Figure 2a shows the current-voltage (*I-V*) curves from a voltage sweep ranging from -3.0 V to +3.0 V, with a 0.01-V interval, measured from 293 K to 363 K (1-K step), limited by our Peltier-driven stage. A nearly-linear characteristics with Ohmic behavior between the bolometric material and the electrode stack was obviously observed. Figure 2b plots the temperature-dependent current values at 3-V derived from the *I-V* curves, demonstrating a significant relationship with the temperature. Figure 2c presents the resistance-temperature (*R-T*) characteristics under constant-voltage mode, which can be modeled following equation to estimate the activation energy (Δ*E*):

$$R(T) = R_o \exp\left(\frac{\Delta E}{k_B T}\right) \quad (2)$$

where *R*(*T*) denotes the temperature-dependent electrical resistance, *T* is the absolute temperature, $R_o$ is a constant, and $k_B$ is the Boltzmann constant. From Equation (2), Δ*E* is extracted from the slope of the Arrhenius plot (ln(*R*) vs. 1000/T) shown in Fig. 2d, which was determined to be 0.315 eV within the measured temperature range. The temperature-dependent current modeling over a high-temperature range, based on the Arrhenius relationship by Equation (2), is described in the inset of Fig. 2b, revealing a rapid, exponential increase in electrical current with rising temperature. However, beyond the certain threshold temperature range (~403 K), our device undergoes degradation due to oxidation-induced alterations in the structural properties of the bolometric material (as detailed in Supplementary Note 6).

To quantify the temperature dependence of the electrical resistance, temperature-coefficient of resistance (TCR), a crucial performance indicator for bolometric detectors, is introduced, which is defined as the derivative of resistance with respect to temperature,

$$\text{TCR} = \left(\frac{1}{R}\right)\left(\frac{dR}{dT}\right) = -\frac{\Delta E}{k_B T^2} \quad (3)$$

Figure 2e shows the temperature-dependent TCR, as obtained by Equation (3). Here, our device achieved a TCR of -4.262 %/K at 293 K, the highest to date for waveguide-integrated PDs utilizing the bolometric effect, thus enhancing bolometric photodetection capabilities.

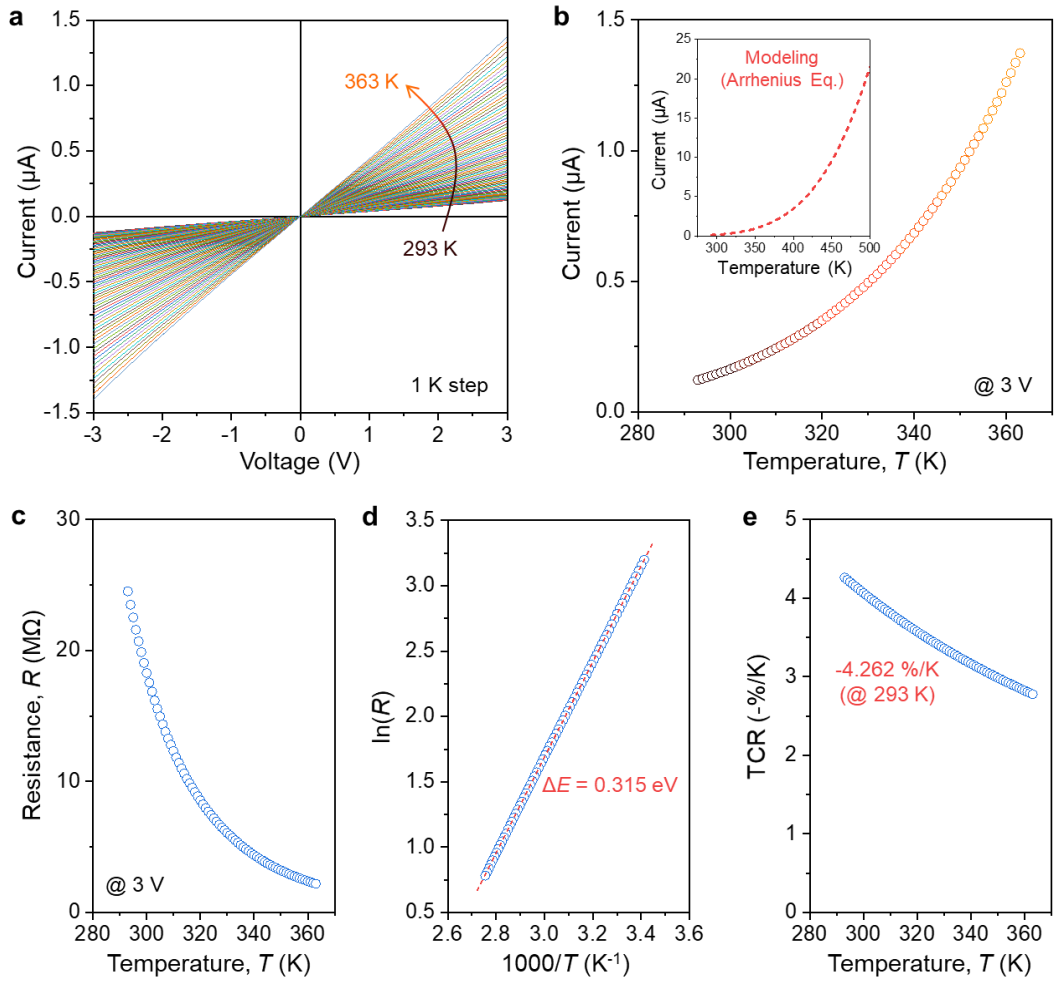

**Figure 2. Thermo-electrical characteristics. a** Current-voltage (*I-V*) curves measured from 293 K to 363 K (1 K step), controlled by a Peltier-driven stage with a 10-mV interval. **b** Temperature dependence of current values at a 3-V biasing. The inset indicates the modeled current values depending on the temperature, calculated from the Arrhenius equation. **c** Resistance-temperature (*R-T*) characteristics under a 3-V biasing, following the Arrhenius relationship. **d** Arrhenius plot [ln(*R*) vs. 1000/*T*] with the activation energy (Δ*E*) calculated from the slope of the linear fit (red dashed line), indicating Δ*E* = 0.315 eV. **e** TCR (-%/K) values depending on the temperature, showing -4.262 %/K at 293 K.

**Photoresponse measurement.** We now turn to explore the MIR photoresponse. The incident optical power coupled into the bolometer region was precisely calibrated, accounting for insertion losses from passive components with the assistance of an identical reference waveguide pattern without the detector part. A total insertion loss of 10.83 ± 0.14 dB (4.18 μm) was used for the calibration (details are provided in Supplementary Fig. S8). The un-illuminated *I-V* curve (off-state), plotted in the inset of Fig. 3a, reveals an off-state current ($I_{off}$) of 127.5 nA at a 3-V bias. Figure 3a presents the change in current, calculated as the ratio of the total measured current ($I_{ph} + I_{off}$) to $I_{off}$, as a function of optical power ($P_{in}$) and the corresponding responsivity (%/mW) at 4.18 μm under a 1 kHz chopping frequency, which is commonly used unit of responsivity for comparing bolometric detectors[26,27,34,35]. We highlight that our device achieved an *R* of 28.77 %/mW (from linear fitting at $P_{in}$ > 0.3 mW), equivalent to voltage responsivity of 863.19 V/W, which is sufficient for a wide range of MIR spectroscopy applications[41–43]. Details for the relationship between responsivity values in different units are described in Supplementary Note 11. This exceeds the previous state-of-the-art values for waveguide-integrated thermal-type PDs beyond 3 μm, which are 24.62 %/mW at 3.8 μm in an Au antenna-assisted PD on an a-Si waveguide using the bolometric effect[27], and 1.97 V/W at 3.7 μm in a graphene-based PD on a GOS waveguide using the PTE effect[29]. A slight nonlinearity is observed at lower $P_{in}$ ranges, potentially attributed to variations in thermo-electrical properties and changes in both $G_{th}$ and $C_{th}$ with temperature. We also estimated the noise-equivalent power (NEP) by taking the ratio of the noise spectral density in the off-state (see Supplementary Note 7) to responsivity at 4.18 μm, calculated as $4.03\times10^{-7}$ W/Hz$^{0.5}$ at 1 kHz. Here, this far exceeds that of previously reported waveguide-integrated MIR PD using the bolometric effect (10.4 μW/Hz$^{0.5}$ at 3.8 μm)[27].

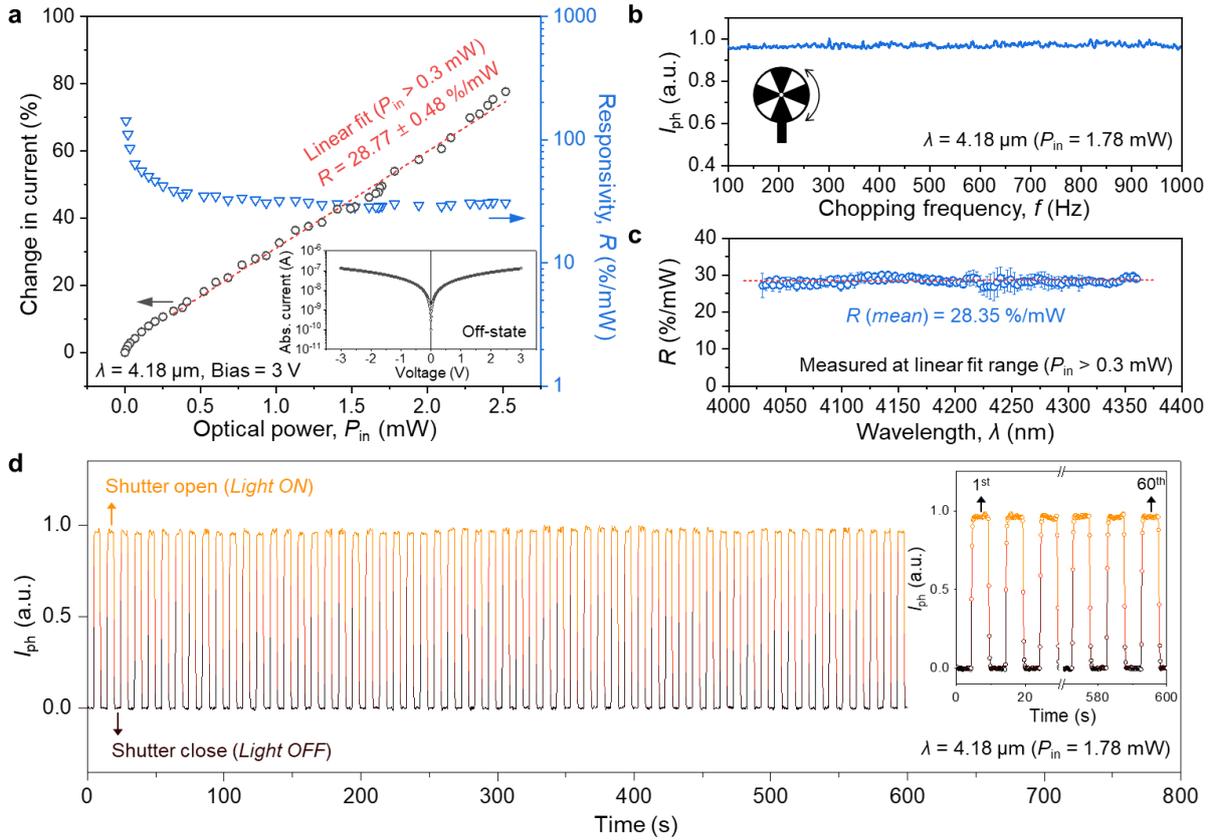

**Figure 3. Photoresponse characterization. a** Change in current (%) as a function of incident optical power ($P_{in}$) toward bolometer region at a wavelength ($\lambda$) of 4.18 μm, showing a responsivity ($R$) of 28.77 %/mW from linear fitting ($P_{in}$ > 0.3 mW). The inset shows the *I-V* curve under off-state condition. **b** Photoresponse under varying chopping frequencies. **c** Spectral response over the range of 4030–4360 nm. **d** Time stability measurement test of photoresponse during continuous ON/OFF cycles with a 10 seconds period, controlled by a beam shutter.

The frequency response was analyzed by varying the chopper frequency. As illustrated in Fig. 3b, our device showed stable performance with a nearly flat response up to 1 kHz (the limit of our setup). Although higher bandwidth might be beneficial, it is not a major concern in most spectroscopy applications, unlike in telecommunications and data communications, and response times of around 1 second are common for many optical gas sensors[44]. This suggests that our device is sufficiently robust for MIR lab-on-a-chip systems, which can be expected to operate at a bandwidth of several tens of kHz level, based on our previous work with the Si-on-insulator (SOI) platform in the near-infrared wavelength range[35]. We also evaluated the spectral response in the MIR band ranging from 4030 to 4360 nm. During the measurement, the $P_{in}$ was maintained within the linear fit region ($P_{in}$ > 0.3 mW). As shown in Fig. 3c, our device exhibited a broadband photoresponse with an $R$ of around 28.35 %/mW across the entire measurable range without any cutoff wavelengths. Lastly, we assessed the long-term stability with switching behavior, a key parameter for evaluating PDs. Notably, as depicted in Fig. 3d, highly stable and repeatable photocurrent generation was observed without noticeable performance degradation throughout the measurements. Here, we note that the response times were constrained by the open/close time of the beam shutter.

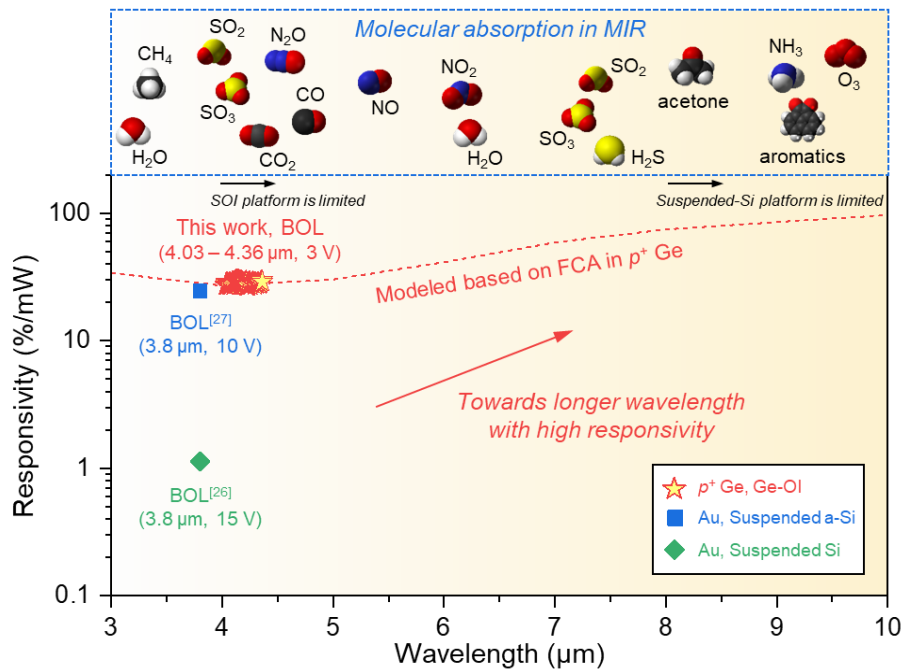

**Figure 4. Benchmark.** Performance comparison of waveguide-integrated PDs using bolometric effect operating beyond 3 μm. The red dotted line represents the responsivity modeling based on FCA in $p^+$ Ge, aligned with our experimental results. Our approach, leveraging the FCA-induced heating mechanism and the Ge-OI platform, can cover the entire MIR spectrum for spectroscopic analysis of numerous biochemical molecules. For each reference, the mechanism of thermal detection, the absorbing medium, and the waveguide platform are indicated. BOL, bolometric. References [26] Wu, Y. et al., *ACS Photonics* **6** (2019); [27] Wu, Y. et al., *Optics Letters* **46** (2021).

We have comprehensively compared our device's performance with that of previously reported MIR waveguide-integrated PDs utilizing the bolometric effect, as shown in Fig. 4. Our device exhibits a broadband responsivity of ~28.35 %/mW (4030–4360 nm), the highest among its counterparts. Notably, to our knowledge, no previous reports have demonstrated waveguide-integrated bolometric PDs operating beyond the wavelength range presented in this work. Traditional Si-based photonics platforms, such as SOI with a limit of ~4 μm and suspended-Si with ~8 μm, inevitably encounter wavelength limitations due to the intrinsic material absorption of $SiO_2$ and Si in the MIR range[45]. Here, by leveraging the FCA-induced heating process on the Ge-OI platform that provides a broad transparency window, our approach can be widely utilized across much shorter or longer wavelength ranges in the MIR spectrum, offering significant potential for spectroscopic analysis of numerous biochemical molecules (shown in Fig. 4) without wavelength constraints. The responsivity modeling, based on FCA in Ge[32] and normalized with our experimental results, is presented in Fig. 4 (red dotted line). Moreover, our device achieves a notable improvement in NEP, over 25 times greater than previously reported waveguide-integrated PDs using the bolometric effect[26,27], without relying on noble metals or exotic materials, thus preserving full CMOS compatibility. Detailed performance characteristics, including comparisons with both bolometric and PTE detectors, are summarized in Supplementary Note 11.

**Sensing demonstration.** To demonstrate the label-free light–analyte interaction capabilities of our MIR PIC-based sensing platform, we arranged a 5-mm-long slot waveguide with our waveguide-integrated PD on a single Ge-OI chip, as shown in Fig. 5a. Efficient mode conversion was facilitated by strip-to-slot and slot-to-strip mode converters[36,38] positioned at the entry and exit points of the slot waveguide (detailed in Supplementary Note 10), respectively, as depicted in the optical microscope and scanning electron microscope (SEM) images in Fig. 5b. The slot waveguide, designed for high confinement within an air-clad, featured geometrical parameters of 1.8 μm width ($W$), 0.2 μm slot gap ($G$), and 500 nm height ($H$), as shown in the cross-sectional SEM image in Fig. 5c, highlighting the well-defined slot region where the light–analyte interaction occurs.

Here, $CO_2$, a major greenhouse gas contributing to global warming[46], was selected as the target analyte with a strong absorption coefficient at 4.23 μm[47]. Under the continuous-wave (CW) operation at 4.23 μm, changes in $CO_2$ gas concentration were detected by the photocurrent signal from our detector while simultaneously monitoring the actual $CO_2$ levels using a commercial gas sensor placed near the device inside the chamber. Operation conditions were maintained at 3 V and 1 kHz for biasing and chopping frequency, respectively. Figure 5d presents the normalized photocurrent signal depending on the $CO_2$ concentration, which exhibits a downward trend as expected from the absorption spectroscopy technique, achieving a sensitivity of 0.0696 %/ppm through linear fitting. Additionally, to assess the repeatability of our optical sensing, we cycled the $CO_2$ valve under nitrogen ($N_2$) gas purging, varying the $CO_2$ concentration between 100 and 250 ppm. As indicated in Fig. 5e, the photocurrent signal varied clearly and repeatably with the $CO_2$ levels, exhibiting no memory effects. It should be noted that the response times were constrained by our experimental setup for both injecting and removing $CO_2$ gas within the chamber.

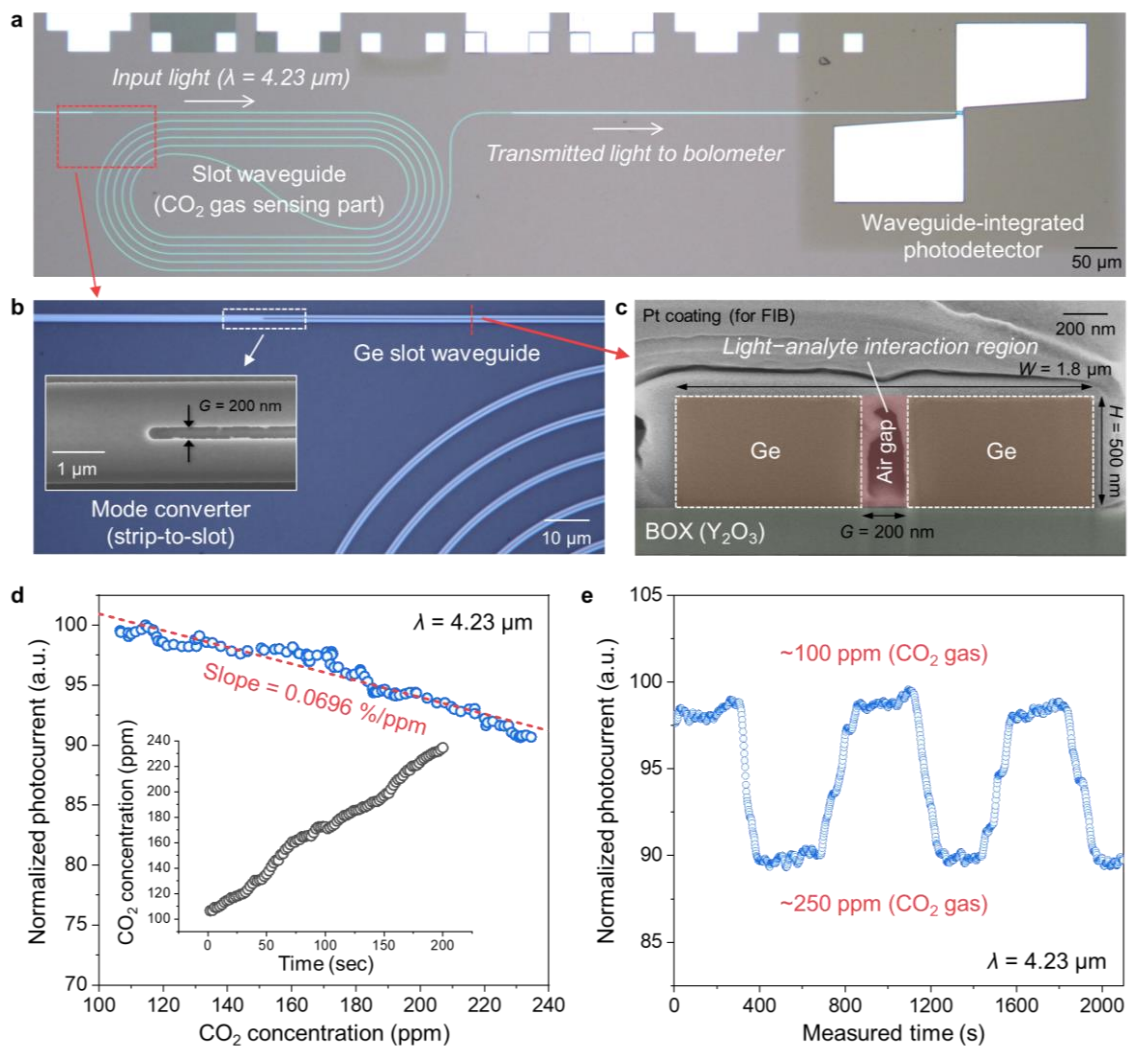

**Figure 5. Optical sensing demonstration. a** Optical microscope image of the integrated photodetector connected with slot waveguide on the Ge-OI platform. **b** Magnified optical microscope image of the Ge slot waveguide. The inset shows the SEM image of the strip-to-slot mode converter. **c** Cross-sectional SEM image showing the slot waveguide with a width ($W$) of 1.8 μm and a gap ($G$) of 200 nm where the light-analyte interaction occurs. **d** Normalized photocurrent at wavelength of 4.23 μm and 3-V biasing, measured under varying $CO_2$ gas concentration. The inset shows the $CO_2$ concentration monitored by a commercial $CO_2$ gas sensor during the photocurrent measurement. **e** Repeatability test of photocurrent at varying $CO_2$ concentrations. All measurements were performed at room temperature.

# Discussion

In the field of integrated photonics, unlike in imaging optics, thermal detection mechanisms have yet to be fully harnessed. As noted in Supplementary Note 11, there have been limited experimental demonstrations to date (bolometric[26,27] and PTE[28,29] effect), primarily due to the challenges of implementing these mechanisms into MIR waveguide platforms. While PTE-based PDs hold promise for uncooled MIR detection with faster response times, a metric less crucial for spectroscopy applications, they face inherent limitations related to the thermoelectric materials. Low-dimensional materials, such as graphene, black phosphorus, and transition metal dichalcogenides (TMDs), are commonly used due to their high Seebeck coefficients[30], but their reliance on exotic materials and lack of CMOS compatibility restrict their widespread adoption. Additionally, the PTE effect—though not bandgap-limited—strongly depends on the material's absorption coefficient, which varies significantly with wavelength, posing challenges for broadband, wavelength-insensitive detection. Moreover, PTE detectors fundamentally rely on the temperature gradient across the material by the Seebeck effect, rather than directly on the amount of light-induced temperature change itself, implying that inconsistent thermal distribution can lead to unpredictable photoresponse. For example, the use of low-dimensional materials, particularly graphene[28,29], exacerbates this issue due to their susceptibility to fabrication tolerances, including surface unevenness and non-uniform doping, which can cause abrupt changes in the Seebeck coefficient. This uncertainty makes PTE detectors less suitable for many MIR spectroscopy applications, especially those requiring stable performance across a broad wavelength range. These challenges not only degrade photodetection capability but also impede the realization of scalable configurations with CMOS integration. Bolometric detectors, in contrast, bypass the inherent limitations of the Seebeck effect by directly converting light-induced temperature variations into changes in electrical resistance, offering a more reliable and practical solution for MIR detection without complex integration hurdles. Thus, we believe that leveraging bolometric effect presents a more attractive strategy for MIR spectroscopy, providing a clear pathway for future advancements in ultra-broadband photodetection and seamless integration with CMOS-compatible platforms.

Here, we present a straightforward, highly CMOS-compatible method for realizing bolometric photodetection within waveguide structures, serving as a blueprint for next-generation MIR PICs. The key idea of our strategy is primarily based on the FCA-induced thermalization process in heavily-doped Ge, providing a viable solution for light-to-heat conversion in MIR absorption without relying on exotic materials in CMOS or hybrid integrations. The ion implantation process is a well-established technique in current semiconductor technologies, enabling fabrication ease, cost-effectiveness, and scalability. Additionally, this approach can be readily adopted into various Si- or Ge-based MIR photonic platforms, including SOI[48], suspended-Si[45], GOS[49], suspended-Ge[50], Ge-on-SOI[51], and Ge-OI[36–39], making it a foundry-friendly solution with the potential of process design kits (PDKs) for large-scale MIR PICs. Furthermore, as bolometry with FCA in Ge is not wavelength-specific, the operational range is extendable across the entire MIR spectrum without encountering cutoff regions.

We have achieved outstanding bolometric photoresponse characteristics through several strategic interventions: enhancing FCA-induced heating in Ge, optimizing device geometries, and improving thermo-electrical properties. Here, increasing the bias voltage may bring into reach further high photocurrent generation up to the breakdown regime at the expense of a rise in the off-state current level (detailed in Supplementary Note 8). Additionally, we can potentially push the photoresponse to a much higher level through various strategies, such as scaling down the width of bolometer region and implementing a top-contact electrode scheme above the dielectric cladding with via plugs[52], improving heat confinement within the absorbing medium. Moreover, thermal-isolation designs, such as air-trench or free-standing structures underneath the $p^+$ Ge region, are advantageous for boosting heating efficiency and mitigating thermal crosstalk between adjacent detectors in an array configuration, albeit at the cost of increased response time. Regarding the bolometric material of the $TiO_2/Ti/TiO_2$ tri-layer film stack, engineering the thickness of each layer and the annealing condition provides the flexibility to tailor electrical resistivity and thermo-electrical properties to meet the requirements of diverse spectroscopy applications[33].

We have successfully demonstrated label-free light–analyte interaction of $CO_2$ molecules using our PIC-based sensing system on the Ge-OI platform. While there are few reports of molecule detection utilizing MIR waveguide-based sensors with monolithic integration of detectors at room temperature[16,53], our work is pioneering in realizing molecule sensing with a CMOS-compatible solution. Moreover, given the ultra-broadband photoresponse characteristics of our approach, we envision the full potential of broad applicability in MIR spectroscopic sensing of various biochemical molecules, as well as real-time detection of multiple analytes, leveraging the label-free nature of the absorption spectroscopy method. Integrating MIR sources, such as interband cascade lasers (ICL)[54] and quantum cascade lasers (QCL)[55], could further enhance our approach, thus paving the way for a fully integrated MIR PIC-based lab-on-a-chip system. Furthermore, employing computational spectroscopy techniques, particularly through disordered structures[9,56,57], could serve as a strategy to minimize system footprint and power consumption while improving robust and agile multiplexed-detection capabilities.

# Materials and methods

**Simulation.** The numerical simulations were conducted using the commercial simulation software packages of Ansys Lumerical, specifically the 3D-FDTD (finite-difference time-domain) and HEAT solvers. The Ge-OI structure was designed with a 500 nm-thick top layer of Ge, a 2 μm-thick $Y_2O_3$ BOX and a Si substrate. For simplicity in analysis, the bolometric material was modeled as a 50 nm-thick $TiO_2$ layer. For the steady-state thermal simulation, the heat source was imported from the absorption data obtained by the 3D-FDTD solver. The background temperature was set at 300 K. A doping concentration in the $p^+$ Ge region was assumed to be $10^{20}$ cm$^{-3}$. Changes in refractive index and absorption coefficient were calculated based on the literature[32] and these values at specific wavelengths were approximated using linear interpolation.

**Device fabrication.** The fabrication process flow of the waveguide-integrated PD using the bolometric effect on the Ge-OI platform is illustrated in Supplementary Note 4. It began with the fabrication of a Ge-OI wafer, which features a 500 nm-thick top Ge, a 2 μm-thick $Y_2O_3$ buried oxide layer, and a Si substrate. We first prepared two types of wafers: (i) the acceptor wafer, which is a Si(100) substrate, and (ii) the donor wafer, comprising a Ge(100)/$Si_{0.5}Ge_{0.5}$/Ge strain relaxed buffer (SRB) layer stack on Si(100) substrate with a thickness of 500 nm, 10 nm, and 900 nm, respectively, grown by the metal-organic chemical vapor deposition (MOCVD) method. We introduced the $Si_{0.5}Ge_{0.5}$ and Ge SRB layers, thereby obtaining a high-quality Ge epitaxial layer with a reduction of lattice mismatch between Si and Ge. We then deposited a 1 μm-thick $Y_2O_3$ layer on both donor and acceptor wafers using the radio-frequency (RF) magnetron sputtering method at 150 °C, ensuring crack-free oxide films. Prior to the direct wafer bonding (DWB) process, we conducted chemical mechanical polishing (CMP) with a silica slurry to planarize the surfaces. As a result, we achieved a surface roughness of approximately 0.6 nm (root-mean-square), as measured by atomic force microscopy (AFM) analysis, which was sufficiently clean and smooth for DWB. After the surface cleaning and $O_2$ plasma treatment, we performed the DWB procedure, followed by the removal of the Si substrate with the sequential processes of mechanical grinding and selective etching using a diluted tetramethylammonium hydroxide (TMAH) solution at 90 °C. The Ge SRB and $Si_{0.5}Ge_{0.5}$ layers were etched away using the APM solution (ammonia hydroxide-hydrogen peroxide water mixture) and the diluted TMAH solution, respectively, resulting in the successful fabrication of the Ge-OI wafer. We then performed electron-beam (e-beam) lithography (NanoBeam Ltd, nB5) with a negative e-beam resist (AR-N 7520) to pattern passive devices; afterwards, an inductively coupled plasma reactive ion etching (ICP-RIE) process (15 sccm $C_4F_8$ and 40 sccm $SF_6$ at a pressure of 25 mTorr, ICP power of 600 W, and RF power of 50 W) was implemented to realize an etching depth of 500 nm, followed by the removal of the e-beam resist in acetone. Prior to ion implantation for forming a $p^+$

Ge region, we deposited a 20 nm-thick dielectric $SiO_2$ layer at 150 °C by atomic layer deposition (ALD) to protect the surface of Ge layer from potential contamination or damage caused by high-energy ions and the removal of photoresist (PR) mask, used for defining the doping region ($p^+$ Ge) while other parts of the sample remained un-implanted. Subsequently, the ion (boron) implantation process was conducted with an implant energy of 110 keV and a dopant dose of $5\times10^{15}$ $cm^{-2}$, followed by activation annealing at 350 °C in an $N_2$ ambient for 3 min. A 25 nm-thick $Al_2O_3$ insulating layer was grown with ALD at 170 °C to eliminate undesirable leakage currents through the Ge layer. A wet etching process using a diluted hydrofluoric acid (HF) solution was then conducted to selectively remove the $SiO_2$ and $Al_2O_3$ oxide layers, while preserving the area surrounding the $p^+$ Ge region. Subsequently, the bolometric material, consisting of a $TiO_2/Ti/TiO_2$ (25/2/25 nm) tri-layer film stack, was sequentially deposited with e-beam evaporation, followed by an acetone lift-off process. Finally, we formed an electrode stack of Ti/W (100/150 nm) with a spacing of ~2 μm through e-beam evaporation and direct current (DC) sputtering, respectively, followed by lift-off using an acetone soak.

**Electrical characterization.** The electrical properties of the fabricated device were characterized using a semiconductor parameter analyzer (Keithley 4200A-SCS) in a four-point probe system equipped with a thermoelectric Peltier-driven stage to precisely control of the background temperature. For DC current-voltage (*I-V*) characterization, the source measure units (SMUs) were utilized to perform a voltage sweep while simultaneously measuring the current with high resolution and accuracy. For low-frequency noise (LFN) analysis, the pulse measure units (PMUs) were employed to capture time-varying current fluctuations, followed by fast Fourier transform (FFT) calculations to analyze the data.

**Photoresponse and gas sensing characterization.** We built an in-house measurement system for photoresponse and gas sensing, as illustrated in Fig. 6. A tunable QCL source (Daylight Solutions MIRcat-QT-2100) under CW operation mode was modulated by a chopper (Scitec Instruments) with a specific reference frequency. Since our grating couplers (GCs) are optimized for the TE-mode[36], we employed a polarizer to maximize optical coupling efficiency. After that, the input light was launched into an indium fluoride ($InF_3$) single-mode optical fiber (Thorlabs) through an aspheric lens with a fiber holder. Alignment between the cleaved $InF_3$ fiber facet and the fabricated device was precisely achieved using a goniometer and three-axis translational stages with a charge-coupled device (CCD) camera. For the characterization of the passive components, the output light was directed into another $InF_3$ optical fiber and subsequently collimated into a mercury cadmium telluride (MCT) detector (VIGO systems PVI-4TE-5). The amount of coupled optical power was calibrated using an external MCT photodiode power sensor (Thorlabs S180C). Propagation losses of strip and slot waveguides were analyzed using the cut-back method, as detailed in Supplementary Note 9. During photoresponse measurements, a programmable current amplifier (Keithley 428-PROG) with two probe contacts was used to apply bias voltage, and the resulting photoresponse signals were monitored on a lock-in amplifier (Stanford Research Systems SR830) for the signal-to-noise (SNR) ratio enhancement. For time stability measurements, a beam shutter (Thorlabs SHB1T) controlled by a function generator (FG, Tektronix AFG3022B) was implemented to turn the coupled light on and off during CW operation. All photoresponse measurements were conducted within an acrylic chamber under a pure $N_2$ gas purging state to mitigate any undesirable impact on $CO_2$ gas absorption from the atmospheric environment. For optical gas sensing demonstrations, $CO_2$ concentrations were regulated using a mass flow meter with a diluting pure $N_2$ gas into the acrylic chamber, which were precisely calibrated by a commercial $CO_2$ gas sensor (Sensirion AG). Here, the lowest achievable $CO_2$ concentration within the acrylic chamber was around 100 ppm, limited by the capabilities of our setup under an $N_2$ gas purging state. All measurements were carried out at room temperature.

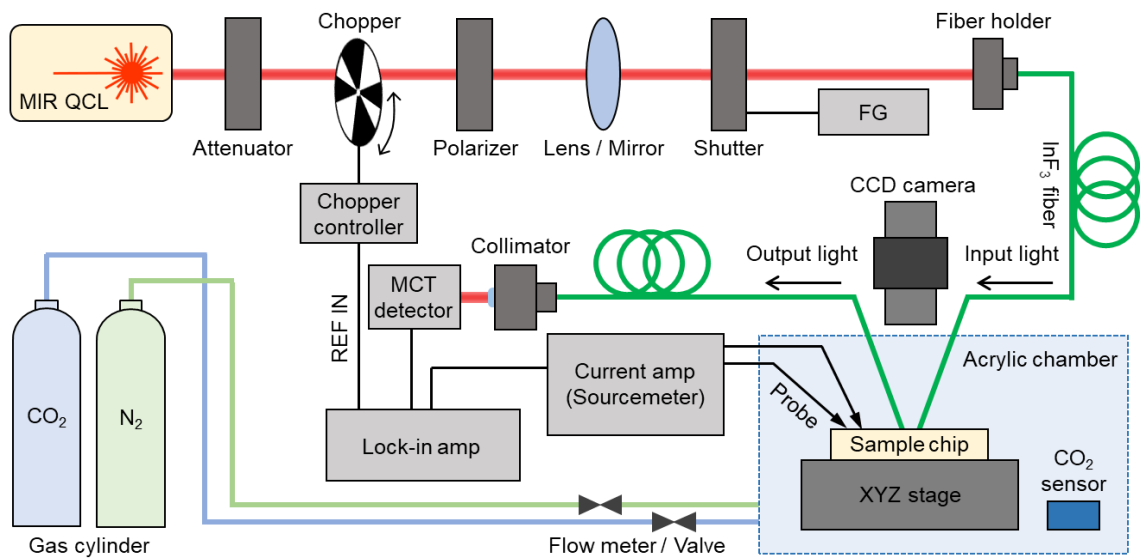

**Figure 6. Schematic of photoresponse and gas sensing measurement setup.** MIR, mid-infrared; QCL, quantum cascade laser; CCD, charge-coupled device, FG, function generator; MCT, mercury cadmium telluride.

# Acknowledgements

This work is supported by the National Research Foundation of Korea (NRF) (2023R1A2C2002777, RS-2024-00407767), the KIST Institutional Program (2E33052), and the BK21 FOUR.

# Author information

**School of Electrical Engineering, Korea Advanced Institute of Science and Technology (KAIST), 291 Daehak-Ro, Yuseong-Gu, Daejeon 34141, Republic of Korea.**

Joonsup Shim, Jinha Lim, Inki Kim, Jaeyong Jeong, Bong Ho Kim, Seong Kwang Kim, and SangHyeon Kim

**Department of Electrical & Computer Engineering, Inha University, 100, Inha-ro, Michuhol-gu, Incheon 22212, Republic of Korea.**

Dae-Myeong Geum

# Corresponding author

Correspondence to SangHyeon Kim.

# Data availability

All data that support the findings of this work are available within the paper. Additional data are available from the corresponding authors upon request.

# Conflict of interests

The authors declare that they have no conflict of interest.

# Author contributions

J. Shim proposed the idea, conducted the numerical simulations, and performed the measurements. J. Shim, J. Lim and I. Kim fabricated the device and built the experimental setup. J. Jeong, B. H. Kim, S. K. Kim, D.-M. Geum, and S. H. Kim contributed to the data analysis and discussion. S. H. Kim supervised the project. All authors contributed to the interpretation of results and prepared the manuscript.

# Supplementary information

# Supplementary Information for

# Room-temperature waveguide-integrated photodetector using bolometric effect for mid-infrared spectroscopy applications

Joonsup Shim[1], Jinha Lim[1], Inki Kim[1], Jaeyong Jeong[1], Bong Ho Kim[1], Seong Kwang Kim[1], Dae-Myeong Geum[2], and SangHyeon Kim[1*]

[1]*School of Electrical Engineering, Korea Advanced Institute of Science and Technology (KAIST), 291 Daehak-Ro, Yuseong-Gu, Daejeon 34141, Republic of Korea*

[2]*Department of Electrical & Computer Engineering, Inha University, 100, Inha-ro, Michuhol-gu, Incheon 22212, Republic of Korea*

*[*]Corresponding Author: SangHyeon Kim (shkim.ee@kaist.ac.kr)*

## Supplementary Note 1. Free-carrier absorption (FCA) and two-photon absorption (TPA) in Ge

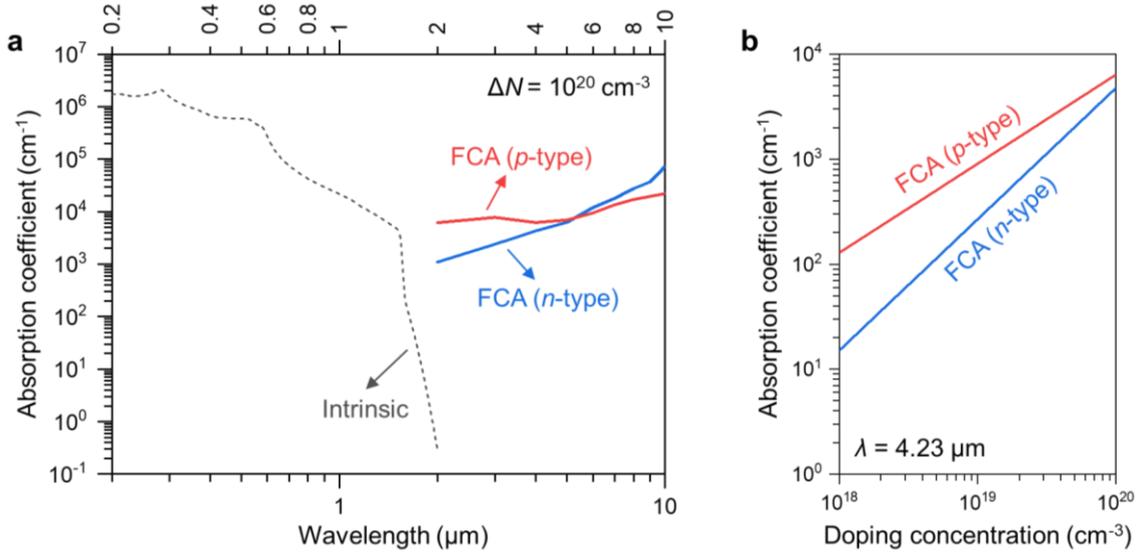

**Fig. S1. FCA in Ge. a** Absorption coefficients in Ge for intrinsic absorption and for FCA in both *n*- and *p*-type Ge with a doping (impurity) concentration ($\Delta N$) of $10^{20}$ cm$^{-3}$ depending on the wavelengths. **b** Comparison of the calculated FCA coefficients as a function of the doping concentration for *n*- and *p*-type Ge at a wavelength ($\lambda$) of 4.23 μm. The absorption coefficients for FCA in Ge are calculated based on the literature[1].

    The free-carrier absorption (FCA) is a process where free carriers, i.e., free electrons and free holes, absorb the incoming photon energy, leading to non-radiative carrier transition with the thermalization process[1,2]. There are two types of FCA in Ge: FCA in *n*-type Ge and FCA in *p*-type Ge, each corresponding to the absorption attributed to the free electrons and free holes, respectively. The absorption coefficients in Ge for intrinsic absorption (band-to-band absorption) and FCA are plotted in Fig. S1a. The doping (impurity) concentration ($\Delta N$) is assumed to be $10^{20}$ cm$^{-3}$. The intrinsic absorption of Ge is negligible beyond the wavelength of ~1.9 μm; however, FCA becomes increasingly significant beyond the bandgap energies. Furthermore, FCA in Ge shows a clear trend of increase with the wavelength of light, highlighting the importance of our approach for broadband mid-infrared (MIR) photodetection – even much higher photoresponse at longer wavelengths – within the waveguide structures on the Ge-on-insulator (Ge-OI) platform. Figure S1b represents the absorption coefficients of FCA for *n*- and *p*-type Ge depending on the doping concentration at the wavelength of 4.23 μm, chosen for the carbon dioxide ($CO_2$) gas sensing demonstration in this work. As shown in Fig. S1b, FCA is more pronounced in *p*-type doped Ge compared to *n*-type. Consequently, we selected *p*-type doping to maximize the FCA-induced heating process in our detector.

    The two-photon absorption (TPA) is a nonlinear optical process where an electron simultaneously absorbs two photons to transition from a lower energy state to a higher energy state. When the photon energy is less than half the bandgap energy, the probability of TPA is significantly reduced. Ge has an indirect bandgap of ~0.66 eV. Since the photon energy in this work (beyond the wavelength of 4 μm) is still below the half of the bandgap energy, the TPA in Ge is significantly reduced or absent. However, TPA can play a role within the doped Ge region since doping introduces impurity energy levels within the bandgap, enabling TPA by providing intermediate states even at photon energies below half the bandgap energy. This can also generate free carriers at a rate proportional to the square of the optical intensity, leading to non-radiative recombination and releasing energy as heat. Here, since the optical power coupled into the bolometer region is much lower than the threshold intensity[3], free-carrier absorption (FCA)-induced thermalization process is the dominant absorption mechanism. However, at intensities

approaching or exceeding this threshold, TPA-induced heat generation can also contribute to the bolometric photoresponse in the proposed detector on the Ge-OI platform.

## Supplementary Note 2. Thermal conductivity analysis of yttrium oxide ($Y_2O_3$) layer

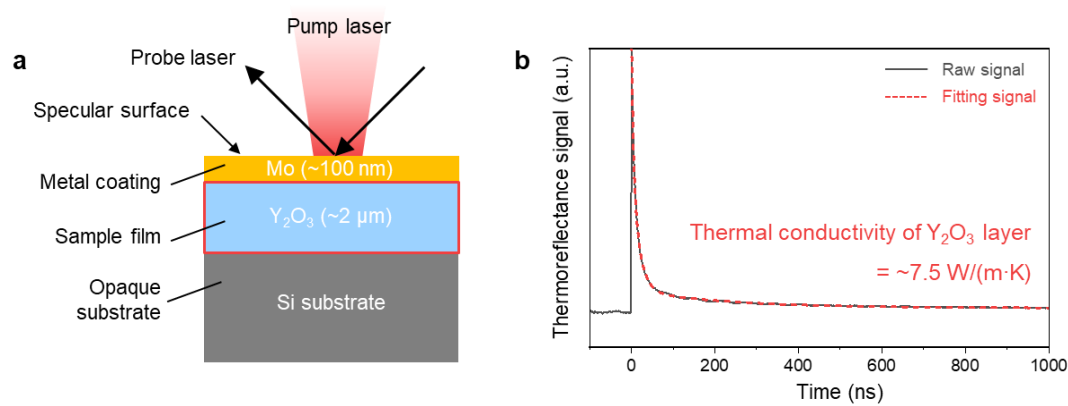

**Fig. S2. Thermal conductivity analysis of an $Y_2O_3$ layer using a time-domain thermo-reflectance (TDTR) front heating/front detection (FF) technique. a** Schematic diagram illustrating the TDTR FF method. **b** Normalized TDTR signal (black line) and the corresponding fitting curve (red dotted line) obtained using the mirror image method. The thermal conductivity of the $Y_2O_3$ layer was determined to be ~7.5 W/(m·K).

To investigate the thermal conductivity of an yttrium oxide ($Y_2O_3$) layer prepared by the radio-frequency magnetron sputtering method in this work, we conducted a time-domain thermoreflectance (TDTR) analysis[4,5] in the front heating/front detection (FF) configuration using a nano-second thermoreflectance apparatus (NanoTR, NETZSCH). This equipment system is a suitable solution for analyzing the thermal properties of thin layers and films, which typically differ from those of bulk materials. The FF measurement configuration can be applied to extract the thermal properties of thin layers on opaque substrates, whereas the rear heating/front detection (RF) method is more appropriate for the sample with transparent substrates.

Figure S2a illustrates a schematic diagram of the sample setup with the TDTR FF technique. The sample consists of a multilayer structure with a 100-nm-thick molybdenum (Mo) layer on top, serving as a transducer, followed by a 2-μm-thick $Y_2O_3$ layer, and a silicon (Si) substrate. The pump laser induces a transient thermal response, which is detected from the front side of the sample. Figure S2b shows the normalized TDTR signal obtained using the FF configuration. The temperature history curve in Fig. S2b reveals an instantaneous temperature rise due to the pump laser irradiation. Subsequently, the temperature gradually decreases as heat diffuses through the Mo layer, eventually reaching the boundary and beginning to diffuse into the $Y_2O_3$ layer. Using the mirror image method[6] (fitting curve), we analyzed the behavior of decaying signal. The thermal conductivity of the $Y_2O_3$ layer was estimated to be ~7.5 W/(m·K).

# Supplementary Note 3. Optimizing device geometries

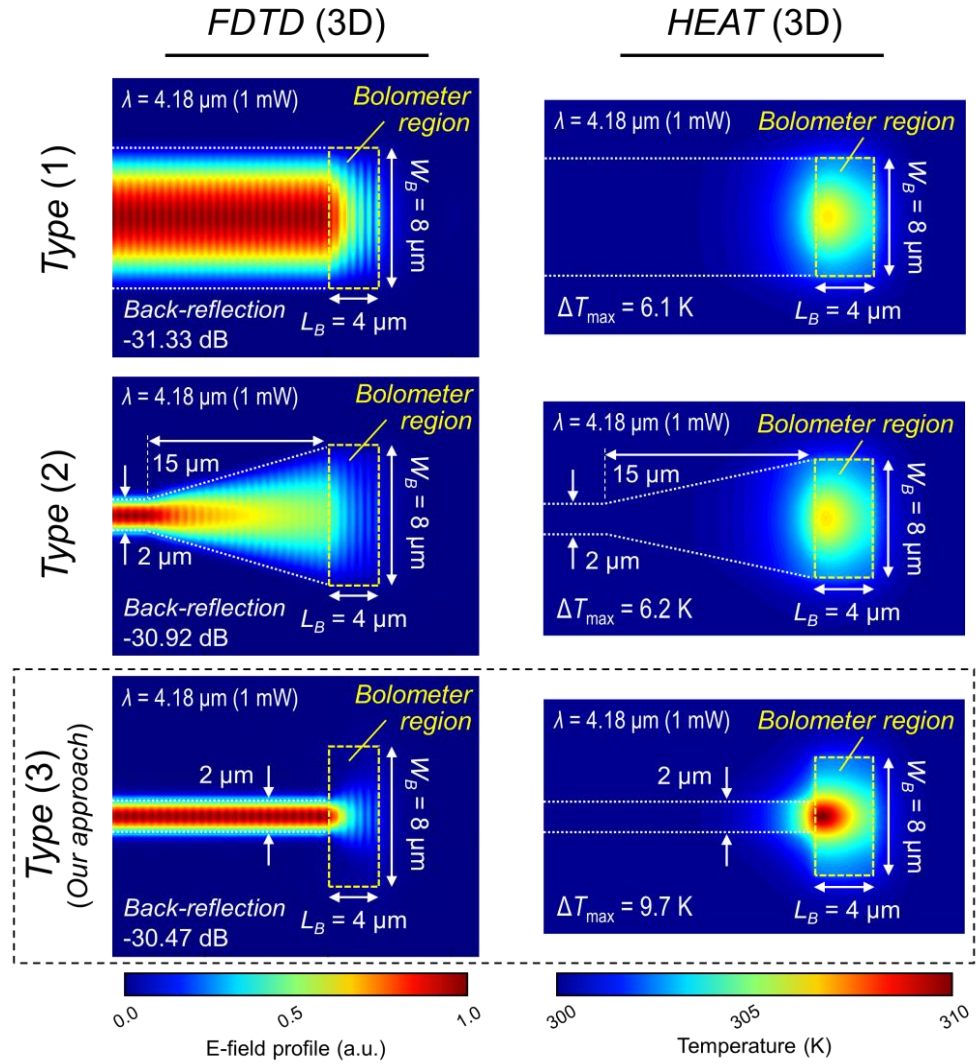

**Fig. S3. Optimization of waveguide-to-bolometer interfacing geometries.** The amount of back-reflection and FCA-induced heating efficiency was obtained through numerical simulations based on different interfacing geometries. Three device configurations were evaluated: Type (1) features a conventional structure, Type (2) incorporates a tapered structure, and Type (3) – our approach – utilizes an abrupt structure.

The optimization process of device geometries was conducted by numerical simulations of 3D-FDTD and HEAT solvers (Ansys Lumerical). We first introduced a terminated-waveguide structure, in which the input waveguide is terminated after the bolometer region ($p^+$ Ge region). Furthermore, we carefully considered the waveguide-to-bolometer interfacing geometry. Figure S3 describes a comparative analysis of three device configurations depending on the interfacing geometries: type (1) – conventional structure, type (2) – tapered structure, and type (3) – abrupt structure (our approach). An input light of 1 mW at 4.18 μm was assumed to support only the fundamental transverse-electric (TE) mode. Here, we primarily considered the amount of back-reflection and the FCA-induced heating efficiency. Type (1) illustrates a conventional waveguide with a uniform width leading directly into the bolometer region, which shows a back-reflection of -31.33 dB and a maximum temperature increase ($\Delta T_{max}$) of 6.1 K. Type (2) depicts a tapered waveguide structure that gradually widens towards the bolometer region, resulting in a back-reflection of -30.92 dB and a $\Delta T_{max}$ of 6.2 K. Type (3) represents an abrupt transition design with a narrow waveguide expanding suddenly to the relatively wider bolometer region, producing a similar lower back-reflection of -30.47 dB and a notably higher $\Delta T_{max}$ of 9.7 K compared with other approaches. Consequently, we opted for the type (3) – abrupt structure – as our approach for enhanced FCA-induced heating performance with reasonable amount of back-reflection.

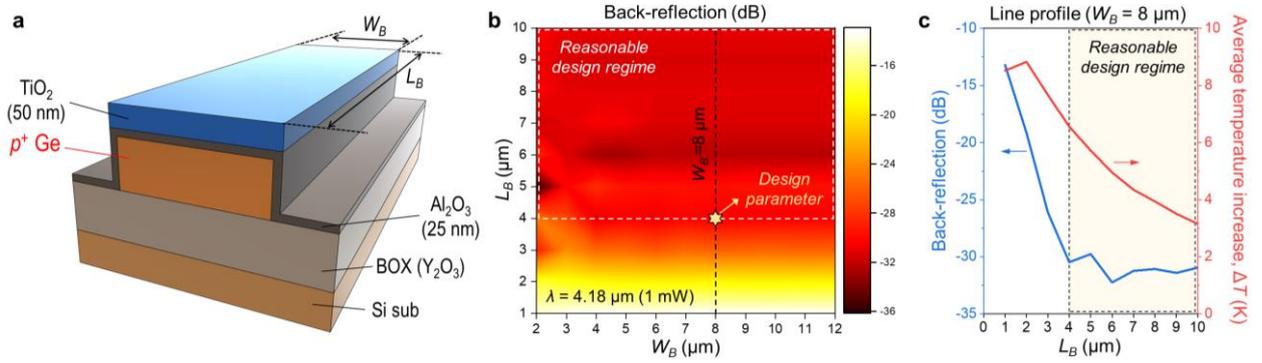

**Fig. S4. Optimization of geometrical parameters of the bolometer region. a** Schematic of the device structure for simulations. **b** Simulated contour map showing geometry-dependent back-reflection across varying $L_B$ and $W_B$ of the bolometer region. **c** Line profiles of both back-reflection and average temperature increase ($\Delta T$) within the bolometer region as $L_B$ varies for a fixed $W_B$ of 8 μm.

Figure S4 presents the simulation results used to determine the optimal geometrical parameters of the bolometer region. A schematic of the simulated device structure is illustrated in Fig. S4a. For these simulations, we explored a wide range of lengths ($L_B$) and widths ($W_B$) for the bolometer region, while the width of an input waveguide ($W_{in}$) leading to the bolometer region was fixed at 2 μm. We first evaluated back-reflection, which can cause unwanted ripples that deteriorate the overall noise characteristics of the sensing system. Figure S4b represents a simulated contour map that shows the geometry-dependent back-reflection. At smaller $L_B$ values, relatively higher back-reflection occurs due to insufficient light absorption in the bolometer region, while changes in $W_B$ have little impact. As shown in Fig. S4b, an $L_B$ greater than 4 μm is preferable, as it ensures back-reflection remains below approximately -30 dB, indicated by a white dotted line and considered within a reasonable design regime. Figure S4c depicts the line profiles of both back-reflection and average temperature increase ($\Delta T$) within the bolometer region, as $L_B$ varies for a $W_B$ of 8 μm. Although smaller $W_B$ could potentially enhance heat confinement, a width of 8 μm was chosen due to alignment tolerance constraints in subsequent in-house fabrication steps. Additionally, $\Delta T$ was calculated within the bolometer region for a central width of 2 μm, corresponding to the spacing of electrodes, due to the minor amount of transfer length between the prepared bolometric material and electrode stack. As shown in Fig. S4c, we finally selected an $L_B$ of 4 μm and a $W_B$ of 8 μm to achieve the highest $\Delta T$ within the reasonable design regime. It is important to note that the geometrical parameters of the bolometer region can be tailored based on the specific application requirements.

# Supplementary Note 4. Fabrication process flow

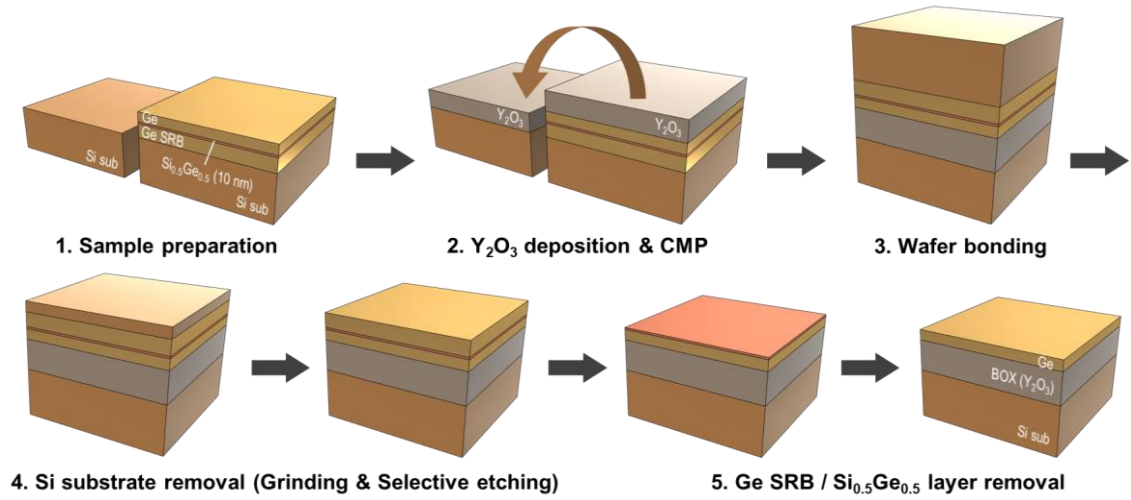

**Fig. S5. Fabrication process flow of the Ge-OI photonic platform using direct wafer bonding (DWB) technique.** SRB, strain relaxed buffer; BOX, buried oxide; CMP, chemical mechanical polishing.

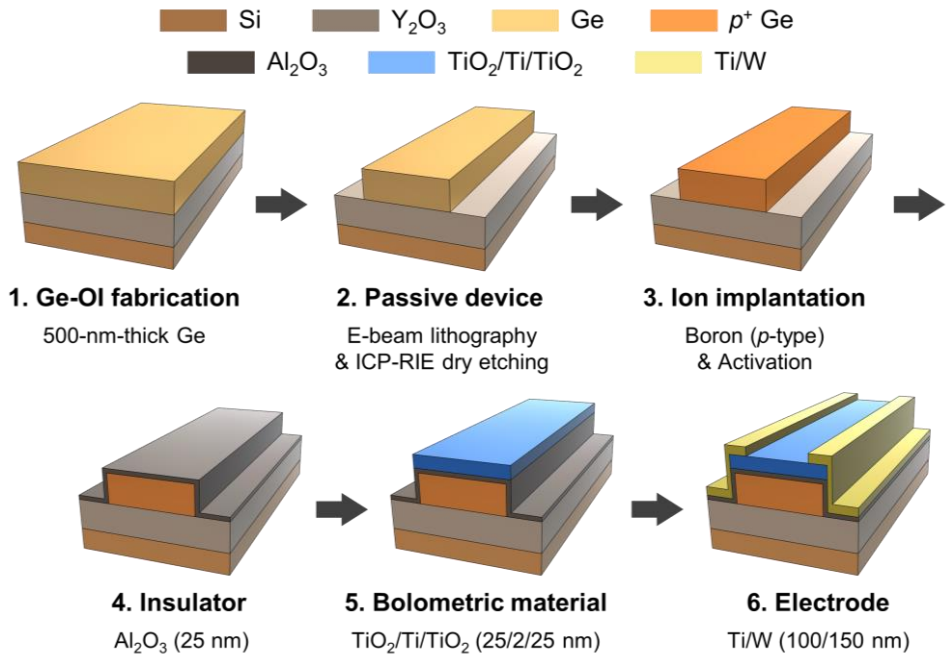

**Fig. S6. Fabrication process flow of the proposed MIR waveguide-integrated PD on a Ge-OI platform.** ICP-RIE, inductively coupled plasma reactive ion etching.

Figures S5 and S6 summarize the fabrication process flow of the proposed MIR waveguide-integrated photodetector (PD) on the Ge-OI platform using the bolometric effect with FCA process. Our fabrication flow can be seamlessly integrated into standard complementary metal-oxide-semiconductor (CMOS) fabrication workflows, enabling large-scale and high-volume manufacturing with cost-effectiveness. The following conditions were considered: (1) Use of CMOS-compatible materials (Group IV-based materials) and avoidance of noble metals, which act as contaminants in CMOS fabs, (2) Avoidance of exotic materials that are unavailable or incompatible with standard CMOS fabrication environments, (3) Standard fabrication steps aligned with CMOS processes, including lithography, etching, doping, and deposition techniques, and (4) Processing temperatures that do not exceed the thermal limits of CMOS devices (typically below 500 ºC). For the proof-of-concept demonstration, we adopted electron-beam lithography; however, this could readily be replaced with conventional photolithography since the critical dimension of our device (~200 nm, for slot waveguide structures) is well within the photolithography resolution limits.

## Supplementary Note 5. Device characterization

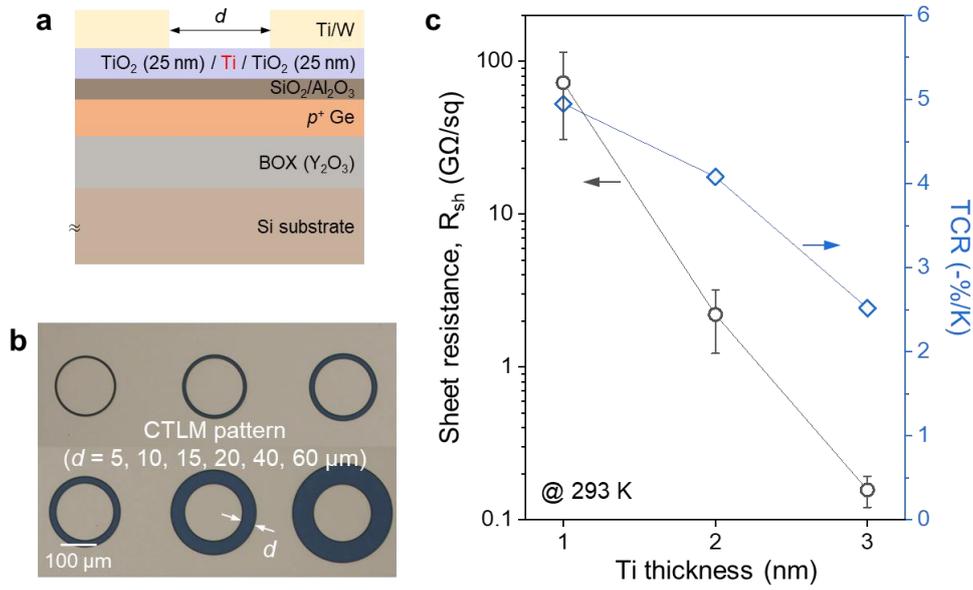

**Fig. S7. Optimization of bolometric material. a** Schematic of the device structure for the CTLM patterns. **b** Optical microscope image of the fabricated CTLM patterns. **c** Variation in sheet resistance (Ω/sq) and TCR (-%/K) depending on the thickness of the metallic Ti layer.

To optimize the thermo-electrical properties of the bolometric material in the $TiO_2/Ti/TiO_2$ tri-layer film, we fabricated the test devices with varying thickness of the metallic Ti layer. Figure S7a illustrates the schematic of these devices, which feature circular transmission line method (CTLM) patterns with spacings (*d*) of 5, 10, 15, 20, 40, and 60 μm, as depicted in the optical microscope image in Fig. S7b. For simplicity, the thickness of the upper and lower $TiO_2$ film was maintained at 25 nm, while the thickness of metallic Ti layer varied from 1 to 3 nm. The thickness of the other layers was kept consistent with those used in our waveguide-integrated photodetector. As demonstrated in Fig. S7c, the extracted sheet resistance ($R_{sh}$) values from the CTLM patterns at 293 K exhibit a clear downward trend as the thickness of the metallic Ti layer increases. Moreover, the temperature-coefficient of resistance (TCR) values, calculated from the temperature-dependent $R_{sh}$ values, show a correlation with the $R_{sh}$ values. Further detailed analysis can be found in our previous work[7]. It is important to note that this approach allows us to tailor the thermo-electrical properties of the bolometric material to meet specific application requirements. Here, we selected a 2 nm thickness for the metallic Ti layer to achieve an optimal balance of electrical resistance and high TCR value in our waveguide-integrated photodetector.

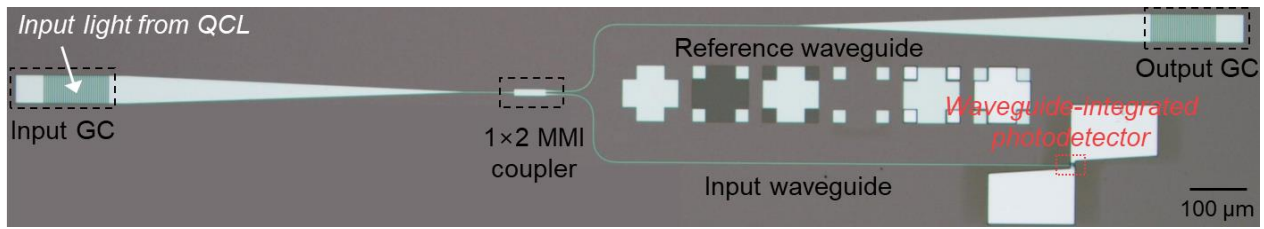

**Fig. S8. Optical microscope image of the fabricated device with the reference waveguide pattern.** QCL, quantum cascade lasers; GC, grating coupler; MMI, multi-mode interference.

Figure S8 shows an optical microscope image of the fabricated waveguide-integrated photodetector with the reference waveguide structure, which was utilized to assist the optical alignment of fiber-optic coupling and to precisely calculate the optical power coupled into the bolometer region. We characterized the coupling efficiencies of our in-house designed grating couplers (GCs)[8] by measuring the insertion loss of GC-to-GC structures with a 1-mm-long channel waveguide between GCs. The coupling loss of each GC was determined as half of the insertion loss, after subtracting the propagation loss of 1-mm-long waveguide (Fig. S15a). Subsequently, we measured the insertion loss of the fabricated device from the input GC to the output GC (Fig. S8) and subtracted the coupling loss of one GC facet (output GC) and the propagation loss of the input waveguide. Accounting for this total insertion loss of 10.83 ± 0.14 dB (4.18 μm), the optical power immediately incident on the photodetector (bolometer region) was calibrated.

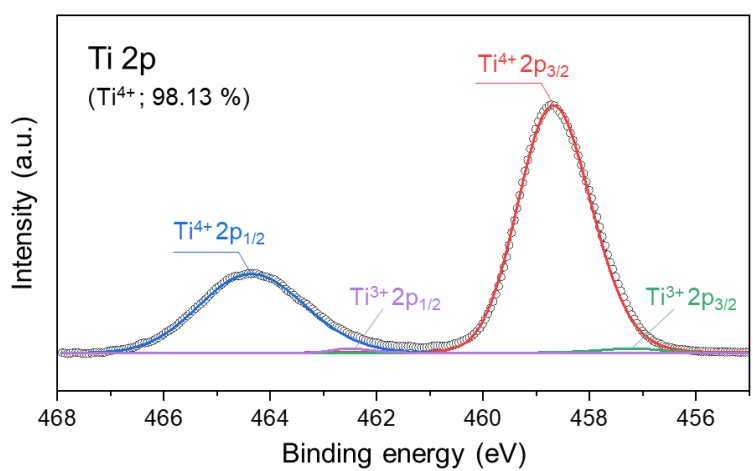

**Fig. S9. XPS analysis.** The 25 nm-thick $TiO_2$ thin film predominantly consists of $Ti^{4+}$ states.

To obtain comprehensive insights into the chemical bonding states and composition of the bolometric material, X-ray photoelectron spectroscopy (XPS) analysis was performed using the Thermo Scientific K-Alpha model at KAIST Analysis Center for Research Advancement (KARA). This analysis employed a monochromatic Al Kα X-ray source to identify the chemical states. Calibration of the energy axis was achieved using the C 1s reference peak. Prior to analysis, the surface of the 25 nm $TiO_2$ thin film was cleaned by sputter-cleaning with an $Ar^+$ ion beam (~5 nm depth) to prevent the effects of surface contaminants on the XPS results. Figure S9 shows the detailed Ti 2p XPS spectra. Deconvolution of these spectra was performed with reference[9] to determine the relative proportions of each valence state. The analysis, as shown in Figure S9, indicated that the $TiO_2$ thin film predominantly consists of $Ti^{4+}$ states (98.13%), with a minor presence of $Ti^{3+}$ states.

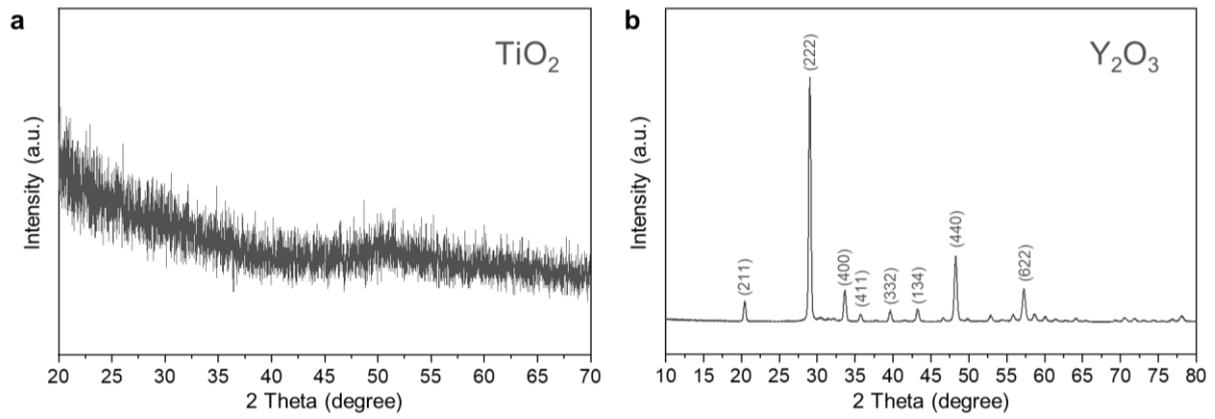

**Fig. S10. XRD analysis. a** XRD analysis result of the $TiO_2$ thin film, showing amorphous state. **b** XRD analysis result of the $Y_2O_3$ layer, showing polycrystalline state.

To investigate the crystalline properties of both the bolometric material (a 25-nm-thick $TiO_2$ film) and the buried oxide (a 2-μm-thick $Y_2O_3$ layer), we performed X-ray diffraction (XRD) analysis using a Cu Kα source in 2-theta scan mode (Rigaku D/MAX-2500 model at KAIST Analysis Center for Research Advancement, KARA). As depicted in Fig. S10a, the diffraction pattern of $TiO_2$ film clearly indicated an amorphous state with no sharp peaks, attributed to the insufficient thermal energy required to achieve crystalline phases. In contrast, as shown in Fig. S10b, the sputter-deposited $Y_2O_3$ layer exhibits clear polycrystalline properties, resulting in its relatively higher thermal conductivity of ~7.5 W/(m·K) (Fig. S2b) compared to typical amorphous state dielectric films[10].

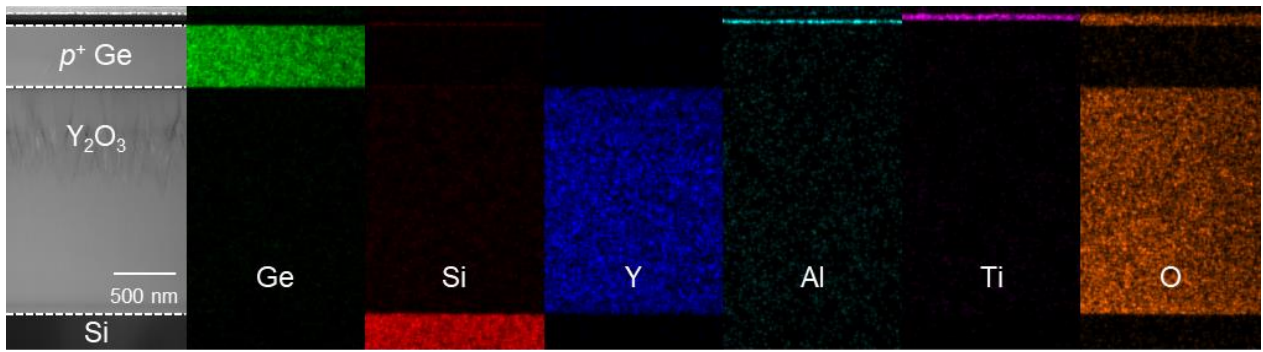

**Fig. S11. TEM with EDS analysis.** Cross-sectional transmission electron microscopy (TEM) image of the fabricated device with energy-dispersive X-ray spectroscopy (EDS) elemental mapping patterns of Ge, Si, Y, Al, Ti, and O atoms.

**Supplementary Note 6. High-temperature stability**

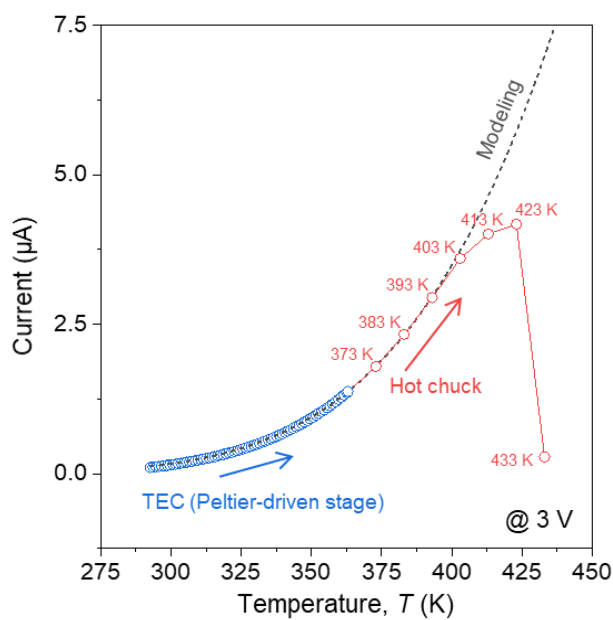

**Fig. S12. High-temperature stability.** Temperature dependence of current values (3 V) measured from 293 K to 433 K, controlled by a Peltier-driven stage (< 363 K) and a hot chuck (> 363 K). The black dotted line indicates modeled current values depending on the temperature, based on the Arrhenius equation.

Figure S12 presents the temperature-dependent current value at a 3 V bias, measured over the range of 293 K to 433 K. The measurements were conducted using a Peltier-driven stage for the lower temperature regime (< 363 K, as shown in Fig. 2b of the main manuscript) and a hot chuck for the higher temperature regime (> 363 K). As shown in Fig. S12, the measured current values closely follow the modeled current (black dotted line) based on the Arrhenius equation up to ~403 K. However, beyond this threshold temperature, the measured current values begin to diverge from the modeled curve. At 433 K, the current level dropped significantly, and the device could not recover its original resistivity, indicating irreversible changes. This deviation from the modeled curve is attributed to the slight oxidation of the metallic Ti layer within the $TiO_2/Ti/TiO_2$ tri-layer film of the bolometric material[7]. Due to the thin Ti layer (~2 nm), it is susceptible to oxidation, even at relatively lower temperatures, leading to a reduction in conductivity. These oxidation-induced changes in the bolometric material at high temperatures (> 403 K) can affect the maximum optical power that can be used with our detector. Based on the relationship between the thermo-electrical properties (Fig. 2b in the main manuscript) and the photoresponse characteristics (Fig. 3a in the main manuscript), we estimate that the incident optical power corresponding to the threshold temperature (~403 K) is ~97 mW, assuming a linear photoresponse. Further optimization of the thickness of each layer in the bolometric material[7] and surface passivation strategies with dielectric materials[11] could enhance high-temperature stability and ultimately elevate the maximum optical power. However, based on our previous work[12], we believe that the relatively low power regime (a few milliwatts) is adequate for achieving a low limit-of-detection (LoD) in optical gas sensing, and further increases in optical power do not result in significant improvements in the LoD. Thus, the estimated maximum optical power of ~97 mW is sufficient for a wide range of MIR spectroscopy applications.

# Supplementary Note 7. Low-frequency noise analysis

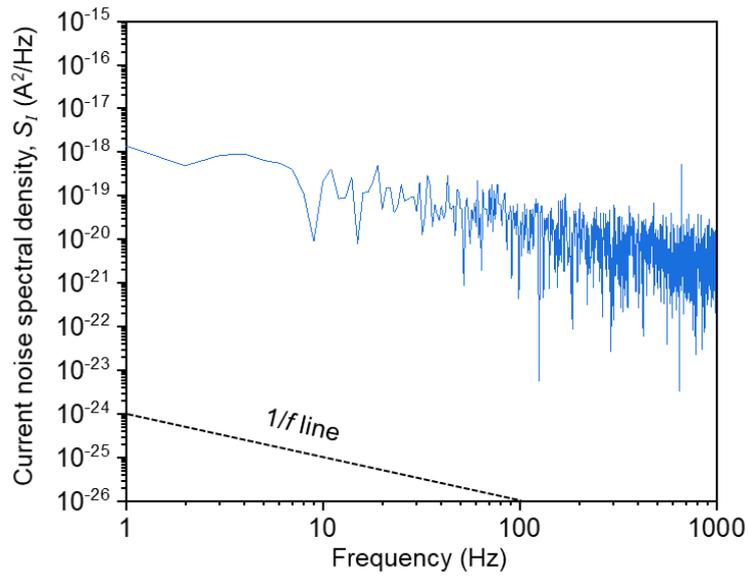

**Fig. S13. Low frequency noise (LFN) characteristics.** Current noise spectral density, $S_I$ (A$^2$/Hz) of the fabricated device as a function of frequency.

Noise characteristics are critical performance metrics for photonic sensing systems. To estimate the noise-equivalent power (NEP) of our device, we examined the low-frequency noise (LFN) features, i.e., current noise spectral density (A$^2$/Hz), without light coupling (dark state). As shown in Fig. S13, the noise characteristic predominantly follows the 1/$f$ line (dashed line), indicating that the measured noise primarily arises from flicker noise, also known as 1/$f$ noise.

# Supplementary Note 8. Electrical breakdown characteristics

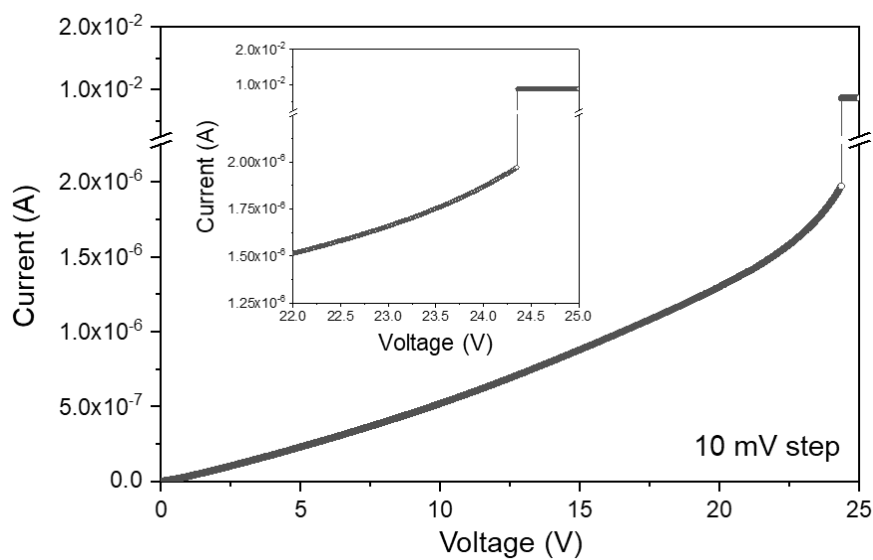

**Fig. S14. Electrical breakdown.** Current-voltage (*I-V*) curve of the fabricated device up to 25 V measured at room temperature with a 10-mV interval, indicating the breakdown voltage of ~23.8 V.

Figure S14 presents the current-voltage (*I-V*) characteristics of our device, which were assessed up to a high voltage range of 25 V at room temperature with a voltage step of 10 mV. It was found that the electrical breakdown occurs at ~23.8 V, characterized by a rapid change in the slope of the *I-V* curve. Beyond this voltage range, the device becomes highly conductive, indicating a transition into the breakdown regime where the device exhibits operational failure.

## Supplementary Note 9. Propagation loss of waveguides

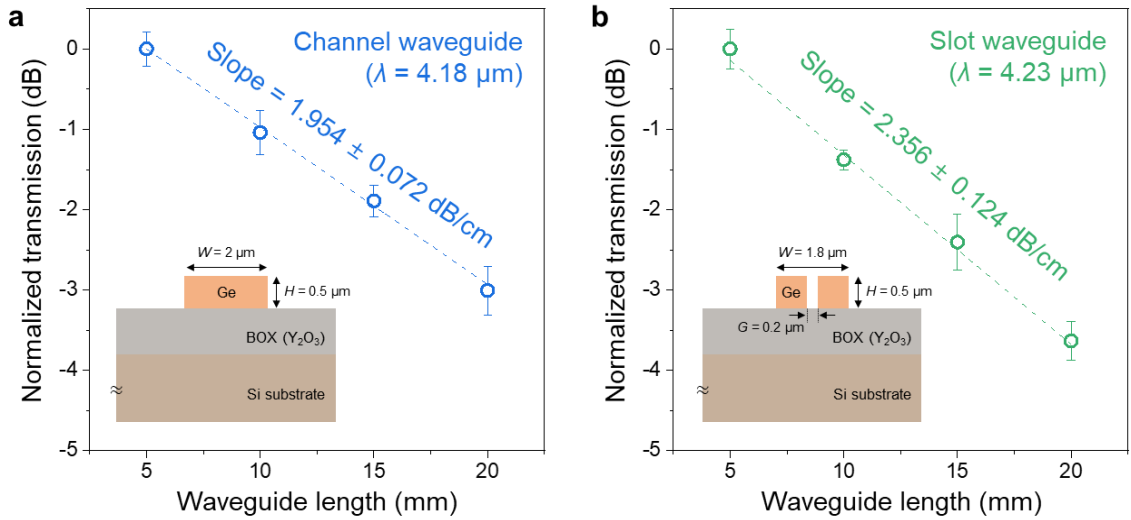

**Fig. S15. Propagation losses. a** Channel waveguide at a wavelength of 4.18 μm. **b** Slot waveguide at a wavelength of 4.23 μm.

Figure S15a and S15b show the propagation losses for channel and slot waveguides on the Ge-OI platform, respectively, as depicted in each inset. These losses were characterized using the cut-back method under an $N_2$ gas purging environment. The results indicate propagation losses of 1.954 ± 0.072 dB/cm for the channel waveguide at a wavelength of 4.18 μm and 2.356 ± 0.124 dB/cm for slot waveguide at 4.23 μm. We found that the propagation losses are relatively large, considering the MIR wavelength range beyond 4 μm. Several strategies can be introduced to reduce the propagation losses: (1) Thermal annealing step with optimized conditions can effectively repair defects in the Ge crystal lattice, such as vacancies and interstitials, caused by electron-beam (e-beam) damage during the e-beam lithography process[13]; (2) Optimization of dry-etching process with $SF_6/C_4F_8$ chemistry, including gas composition, pressure, and power settings, can achieve smoother sidewalls, thereby mitigating the scattering losses[12].

# Supplementary Note 10. Simulation of mode converters

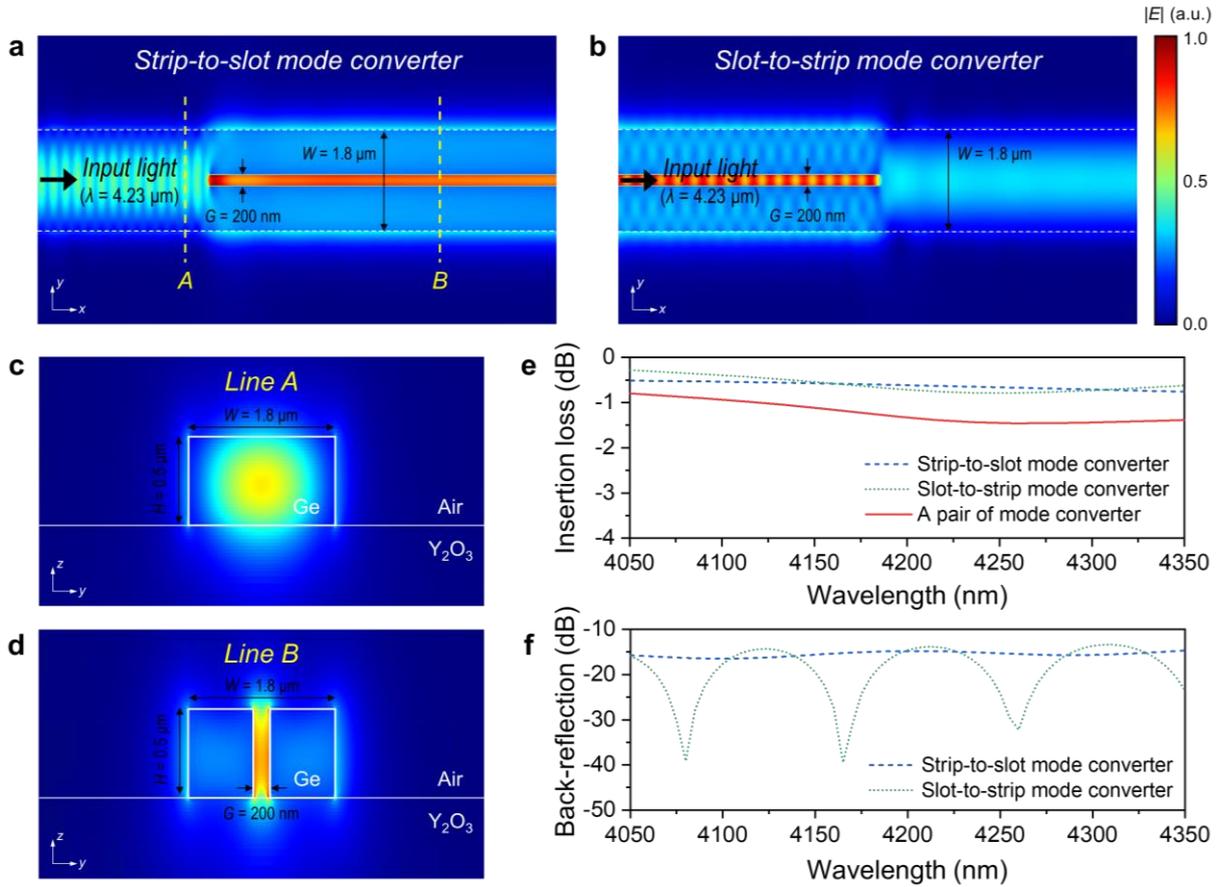

**Fig. S16. Mode converters on Ge-OI platform. a-b** Electric-field distribution for strip-to-slot and slot-to-strip mode converters at a wavelength of 4.23 µm. **c-d** Cross-sectional field distribution along lines *A* and *B* for the channel and slot waveguide regions, respectively. **e** Simulated mode conversion efficiencies for individual converters and a pair of mode converter. **f** Simulated back-reflection of each mode converter.

Figures S16a and S16b depict the electric-field distribution of strip-to-slot and slot-to-strip mode converters on the Ge-OI platform at a wavelength of 4.23 µm in fundamental transverse-electric (TE) mode, respectively. The conversion efficiencies between strip and slot modes[8] were calculated by a 3D-FDTD solver (Ansys Lumerical). Figure S16c and S16d show the cross-sectional field distribution along lines *A* and *B* in the strip and slot regions, respectively, representing well-confined modal fields. The simulated mode conversion efficiencies for the wavelength range from 4050 to 4350 nm are plotted in Fig. S16e. The insertion of a pair of mode converter at 4.23 µm was estimated to be 1.43 dB. Figure S16f shows the back-reflection of the designed strip-to-slot and slot-to-strip mode converter over the wavelength range from 4050 to 4350 nm, with the values of -14.99 dB and -15.51 dB at 4.23 µm, respectively. Here, multiple back-reflections from the converter facet could result in etalon fringe patterns at the transmission spectrum, which can be further suppressed by carefully designing the mode coupling structure[14,15].

# Supplementary Note 11. Performance comparison

**Table S1. The overall performance characteristics of the reported MIR waveguide-integrated thermal-type PDs.** ChG, chalcogenide glass; BOL, bolometric; PTE, photothermoelectric; TCR, temperature-coefficient of resistance; NEP, noise-equivalent power.

| Reference, Year | Ref [16], 2019 | Ref [17], 2021 | Ref [18], 2022 | Ref [19], 2024 | This work |
|---|---|---|---|---|---|
| Operation wavelength | 3.72 – 3.88 μm | 3.72 – 3.88 μm | 5.2 μm | 3.61 – 3.7 μm | 4.03 – 4.36 μm |
| Photonic platform | Suspended-Si (c-Si) | Suspended-Si (a-Si) | ChG-on-CaF$_2$ (Ge$_{28}$Sb$_{12}$Se$_{60}$-on-CaF$_2$) | Ge-on-Si (GOS) | Ge-OI (Ge-on-insulator) |
| Operation mechanism | BOL | BOL | PTE | PTE | BOL |
| Absorption material | Au | Au | Graphene | Graphene | $p^+$ Ge |
| Bolometric material | a-Si | a-Si | - | - | TiO$_2$/Ti/TiO$_2$ |
| TCR (-%/K) | 0.9 | 1.9 | - | - | 4.262 (@293 K) |
| Responsivity (%/mW) | 1.13 (@15 V, 3.8 μm) | 24.62 (@10 V, 3.8 μm) | - | - | 28.77 (@3 V, 4.18 μm) |
| Responsivity (mA/W) | 2.26×10$^{-4}$ (@15 V, 3.8 μm) | 2.95×10$^{-3}$ (@10 V, 3.8 μm) | - | - | 3.669×10$^{-2}$ (@3 V, 4.18 μm) |
| Responsivity, $R$ (V/W) | 169.5 (@15 V, 3.8 μm) | 2458.33 (@10 V, 3.8 μm) | 1.5 (@0 V, 5.2 μm) | 1.97 (@0 V, 3.7 μm) | 863.19 (@3 V, 4.18 μm) |
| Off-state (dark) current (nA) | 20 (@ 5 V) | 12 (@10 V) | Not stated | Not stated | 127.5 (@3 V) |
| NEP (W/Hz$^{0.5}$) | 6.6×10$^{-5}$ | 1.04×10$^{-5}$ | 1.1×10$^{-9}$ | 2.8×10$^{-9}$ | 4.03×10$^{-7}$ |
| $R$ / NEP | 2.57×10$^6$ | 2.36×10$^8$ | 1.36×10$^9$ | 7.04×10$^8$ | 2.14×10$^9$ |
| CMOS-compatibility | Low | Low | Low | Low | High |

As discussed in this work, a bolometer is a class of thermal-type photodetectors that absorbs incident light, leading to an increase in its temperature. This temperature rise results in a change in the bolometer's electrical resistance, due to the strong temperature dependence of its resistive material (bolometric material). The change in resistance serves as the primary signal for determining the intensity of incident optical power. There are two main strategies for detecting change in the electrical resistance of the bolometric material: (1) constant voltage source and (2) constant current source, which measure the current and voltage signals, respectively.

In this work, we adopted the constant voltage source method to measure changes in the electrical resistance of the bolometric material. This approach allowed us to measure the current variation corresponding to varying incident optical power. Here, we introduced the responsivity in terms of '%/mW' to represent the performance of bolometric photodetection, which indicates the percentage change in current as a function of incident optical power. This unit provides a fairer basis for comparison, particularly for bolometric detectors. The conventional unit (A/W or V/W) can vary significantly depending on the electrical resistivity of the bolometric material (resistive film) and the magnitude of operating source (voltage or current), which may lead to misunderstandings of the detector's performance. However, for a comprehensive comparison with previous literatures, as shown in Table S1, we utilized the relationship between responsivity values expressed in different units. Details of these calculations are described below.

## Derivation Steps

The definition of variables used for this derivation steps are below:

*S* [%/mW]: Percentage responsivity in current (percentage change of current as a function of optical power)
$I_{off}$ [A]: Off-state current (dark current)
$R_I$ [A/W]: Current responsivity
$R_V$ [V/W]: Voltage responsivity
$P_{in}$ [W]: Incident optical power
*R* [Ω]: Electrical resistance
*V* [V]: Voltage
*I* [A]: Current

We aim to derive the relationship between the voltage responsivity ($R_V$) and current responsivity ($R_I$) expressed as in Eq. (1):

$$R_V \text{ [V/W]} = -R \text{ [Ω]} \times R_I \text{ [A/W]} \qquad (4)$$

The voltage change (Δ*V*) under constant current bias ($I_{bias}$) can be described as:

$$\Delta V = I_{bias} \times \Delta R \qquad (5)$$

And, the current change (Δ*I*) under the constant voltage bias ($V_{bias}$) is:

$$\Delta I = -\frac{V_{bias}}{R^2} \times \Delta R \qquad (6)$$

(Detailed derivation of Eq. (3) is provided separately below.)

From the definition of current responsivity ($R_I$) and Eq. (3):

$$R_I = \frac{\Delta I}{P_{in}} = -\frac{V_{bias}}{R^2} \times \frac{\Delta R}{P_{in}} \qquad (7)$$

From Eq. (4):

$$\frac{\Delta R}{P_{in}} = -\frac{R^2}{V_{bias}} \times R_I \qquad (8)$$

Substituting back into the definition for $R_V$ and from Eqs. (2) and (5):

$$R_V = \frac{\Delta V}{P_{in}} = I_{bias} \times \frac{\Delta R}{P_{in}} = I_{bias} \times \left(-\frac{R^2}{V_{bias}} \times R_I\right) \qquad (9)$$

Simplify Eq. (6):

$$R_V = -\frac{I_{bias} R^2}{V_{bias}} \times R_I \qquad (10)$$

With the relationship for $I_{bias}$:

$$I_{bias} = \frac{V_{bias}}{R} \qquad (11)$$

From Eqs. (7) and (8), we finally have:

$$R_V = -R \times R_I \qquad (12)$$

In addition, from the definition for current responsivity ($R_I$), the relationship between $R_I$ and $S$ is expressed as:

$$R_I = \frac{\Delta I}{P_{in}} = I_{off}\left(1 + \frac{S}{100} \times 1000\right) - I_{off} = I_{off} \times 10 \times S \tag{13}$$

Consequently, we can obtain the relationship of the Eq. (11) from Eqs. (9) and (10):

$$R_V \text{ [V/W]} = -R \text{ [}\Omega\text{]} \times I_{off} \text{ [A]} \times 10 \times S \text{ [\%/mW]} \tag{14}$$

*Additional Derivation Steps*

Here, we aim to derive the Eq. (12):

$$\Delta I = -\frac{V_{bias}}{R^2} \times \Delta R \tag{15}$$

Under constant voltage bias, the current through our bolometric photodetector is given by *Ohm's Law*:

$$I = \frac{V_{bias}}{R} \tag{16}$$

After the resistance change ($\Delta R$) due to light injection, the current becomes:

$$I + \Delta I = \frac{V_{bias}}{R + \Delta R} \tag{17}$$

$$\Delta I = \left(\frac{V_{bias}}{R + \Delta R}\right) - \left(\frac{V_{bias}}{R}\right) \tag{18}$$

Simplify this expression:

$$\Delta I = V_{bias}\left(\frac{1}{R + \Delta R} - \frac{1}{R}\right) \tag{19}$$

Here, we can consider *I* as a function of *R*:

$$I(R) = \frac{V_{bias}}{R} \tag{20}$$

Differentiating *I* with respect to *R*:

$$\frac{dI}{dR} = -\frac{V_{bias}}{R^2} \tag{21}$$

For small changes, $\Delta I$ can be expressed as:

$$\Delta I \approx \frac{dI}{dR}\Delta R \tag{22}$$

Thus, from Eqs. (18) and (19) we can finally get the Eq. (20), which aligns with Eq (3):

$$\Delta I = -\frac{V_{bias}}{R^2} \times \Delta R \tag{23}$$